\documentclass[9pt]{article}
\usepackage[margin=2cm]{geometry}
\usepackage{authblk}
\usepackage{graphicx}
\usepackage{color}
\usepackage{amsmath}
\usepackage{paralist}
\usepackage{indentfirst}  %% Text feature after captions
\usepackage{bm}    %%%%%  for mathematical symbols
\usepackage{amsfonts}
\usepackage{tikz}
\usetikzlibrary{calc}
\usepackage{adjustbox}
\usepackage{ amssymb }
\usepackage[skins]{tcolorbox}
\usepackage{mathtools}
\usepackage{algorithm}% http://ctan.org/pkg/algorithm
\usepackage{algpseudocode}% http:

\DeclareMathOperator{\arctantwo}{arctan2}
\newcommand{\Il}{\mathcal{I}}
\newcommand{\loss}{\mathcal{H}}
\newcommand{\Pj}{\textbf{P}}
\newcommand{\R}{\mathbb{R}}
\newcommand{\E}{\mathbb{E}}
\newcommand{\Z}{\mathbb{Z}}
\DeclareMathOperator{\wrap}{wrap}
\DeclareMathOperator{\abs}{abs}
\DeclareMathOperator*{\argmax}{arg\,max}
\usepackage{siunitx}

\usepackage{xcolor}

\newcommand{\kinectTM}{Kinect\texttrademark}

\newcommand*{\QED}{\hfill\ensuremath{\square}}%

\usepackage{soul}

\newcommand{\snapshotrotation}[5]{%
\begin{adjustbox}{width=0.98\textwidth}

	\begin{tikzpicture}
	
	\node[anchor=north west,inner sep=0] (img) at (0,0) {\includegraphics{#1}};
	
	 		\node [anchor = north west, scale=2.65] at (0.2, -0.2){(a)};
			
			\node [anchor = north west, scale=2.65] at (#2-0.3-1.75, -0.2){(b)};
  
	  		\draw[draw=black, opacity=1, line width = 0.5 mm]
 	      		(#2-1.75,#3-1.75) -- (#2-1.75,{#3-3.5-1.75}) -- ({#2+1.75},{#3-3.5-1.75}) -- ({#2+1.75},#3-1.75) -- cycle;
	
			\node [below left, text width=1.5cm, align=right, scale=1.5] at ({#2 + 1.75+0.95-1.75},{#3 -3.5-1.75}){$\theta= #5^0$};
		
			\draw[->, line width = 0.25 mm] ({#2+-1.75}, {#3 - 3.5})--({#2 + 3.45-1.75},  {#3 - 3.5}) node[right, scale=2]{$x$};
		
			\draw[->, line width = 0.25 mm] ({#2}, {#3 - 3.5-1.75})--({#2}, {#3 - 1.8}) node[above, scale=2]{$y$};			
			
			\coordinate (centre) at ({#2}, {#3 - 3.5});
		
			\coordinate (arrow1) at ({3.5*.25*cos(-#5)}, {3.5*-.25*sin(-#5)});
			\coordinate (arrow2) at ({3.5*-.25*cos(-#5)}, {3.5*.25*sin(-#5)});
			\draw [ blue, <-> , line width = 0.5mm]  let \p{centre} = (centre), \p{arrow1} = (arrow1), \p{arrow2} = (arrow2) in
				({\x{centre} + \x{arrow1}},
 				{\y{centre} + \y{arrow1}})
							-- 
				({\x{centre} + \x{arrow2}},
 				{\y{centre} + \y{arrow2}});	
			
			\draw [blue,thick,domain=180:180+#5, line width = 0.3mm] let  \p{centre} = (centre) in plot ({#2 - 3.5*0.15*cos(\x)}, {#3 - 3.5 - 3.5*0.15*sin(\x)});
		
			\node [below left, text width=0cm, align=center, scale=2] at (#2 + 3.5*0.83-1.75, #3 - 3.5*0.5 + 0.8 -1.75){$\theta$};
 
	 	    \end{tikzpicture}
\end{adjustbox}
}

\newcommand{\figandor}[2]{%
\begin{tikzpicture}
\draw (0, 4.5) node[anchor=north west,inner sep=0] (img) {#1};
\node [below,text width=0.4cm, align=center, scale=0.75] at (.25, .8){(#2)};
\end{tikzpicture}%
}

\newcommand{\figmerge}[7]{%
\begin{adjustbox}{height=0.163\textheight}
\begin{tikzpicture}
	\draw (0, 0) node[anchor=north west,inner sep=0] (figure) {\includegraphics[height=4.0cm]{#1}};
	\begin{scope}[x={(figure.north east)},y={(figure.south west)}]
	\node [below left, align=center, scale=0.75] at (0.14, 0.86){(#3)};
	\node[anchor=north west, inner sep = 0](trajectory) at (.94,#7) {\includegraphics[height=#4]{#2}};
	\begin{scope}[x={(figure.north east)},y={(figure.south west)}]
	
	\draw[->] (0.965, 0.758)--(1.017,  0.758) node[right, scale = 0.6]{#5};
	\draw[->] (0.965, 0.758)--(0.965,  0.703) node[above, scale=0.6]{#6};
	
	\end{scope}
	\end{scope} 
\end{tikzpicture}
\end{adjustbox}
}

\title{Pedestrian orientation dynamics from high-fidelity measurements}

\author[a]{Joris Willems}
\author[a]{Alessandro Corbetta} 
\author[b]{Vlado Menkovski}
\author[c]{Federico Toschi}

\affil[a]{\small{Department of Applied Physics, Eindhoven University of Technology,  5600 MB Eindhoven, The Netherlands}}
\affil[b]{\small{Department of Mathematics and Computer Science, Eindhoven University of Technology,  5600 MB Eindhoven, The Netherlands}}
\affil[c]{\small{Department of Applied Physics, Department of Mathematics and Computer Science, Eindhoven University of
Technology - 5600 MB Eindhoven, The Netherlands and CNR-IAC, I-00185 Rome, Italy}}

\date{}

\begin{document}

\maketitle

\begin{abstract}
%
% AUTOCOUNT-START wc: 250
  We investigate in real-life conditions and with very high accuracy
  the dynamics of body rotation, or yawing, of walking pedestrians -
  an highly complex task due to the wide variety in shapes, postures
  and walking gestures.  We propose a novel measurement method based
  on a deep neural architecture that we train on the basis of generic
  physical properties of the motion of pedestrians. Specifically, we
  leverage on the strong statistical correlation between individual
  velocity and body orientation: the velocity direction is typically
  orthogonal with respect to the shoulder line. We make the reasonable
  assumption that this approximation, although instantaneously
  slightly imperfect, is correct on average. This enables us to use
  velocity data as training labels for a highly-accurate
  point-estimator of individual orientation, that we can train with no
  dedicated annotation labor.  We discuss the measurement
  accuracy and show the error scaling, both on synthetic and real-life
  data: we show that our method is capable of estimating orientation
  with an error as low as $7.5$ degrees.  This tool opens up new
  possibilities in the studies of human crowd dynamics where
  orientation is key.  By analyzing the dynamics of body rotation in
  real-life conditions, we show that the instantaneous velocity
  direction can be described by the combination of orientation and a
  random delay, where randomness is provided by an Ornstein-Uhlenbeck
  process centered on an average delay of $100\,$ms. Quantifying these
  dynamics could have only been possible thanks to a tool as precise
  as that proposed.
% AUTOCOUNT-END
\end{abstract}

\section*{Introduction}
The orientation of our body and shoulder-line changes continuously as
we walk.  When our gait is regular, these changes are nearly periodic
and follow the swinging trend of our trajectories as we balance our
weight between our feet~\cite{pontzer2009control}.
% paced by the shift of body weight between our feet.
At times, motion direction and body orientation remain temporarily
decoupled. This happens, for instance, when we sidestep or in
proximity of turns and distractions.

Shoulder-line yawing is not just a mechanical reflection
of the walking action, it rather becomes an essential dynamic
ingredient as our motion gets geometrically constrained, e.g. by a
dense crowd or by a narrow environment. In both cases, as we need to
make our way to our destination, we, consciously or unconsciously,
rotate our bodies sideways to minimize collisions or maintain comfort
distances with other pedestrians or the environment.

\begin{figure} [ht!]
\centering
\begin{tikzpicture}
\draw (0, 0) node[anchor=north west,inner sep=0] (img) {\includegraphics[width=0.38\textwidth]{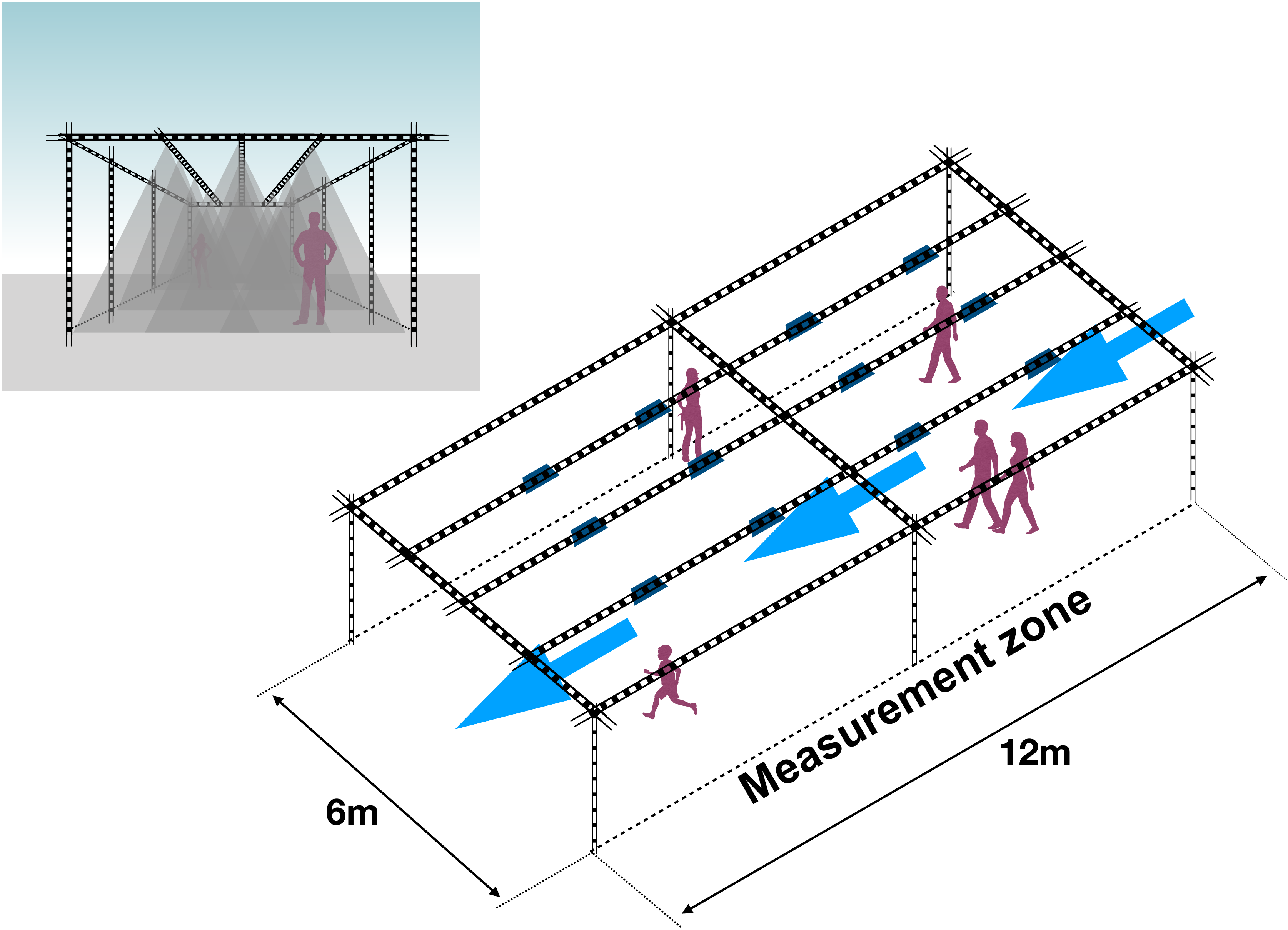}};
\begin{scope}[x={(img.north east)},y={(img.south west)}]
	\node [anchor = north west, text width=0.25cm, align=center, scale=1.2] at (0, 0){(a)};
	\node [anchor = north west, text width=0.25cm, align=center, scale=1.2] at (0.68, 0.0){(b)};
\end{scope}
\end{tikzpicture}
\begin{tikzpicture}
\draw (0, 0) node[anchor=north west,inner sep=0] (img) {\includegraphics[width=0.28\textwidth, angle=90]{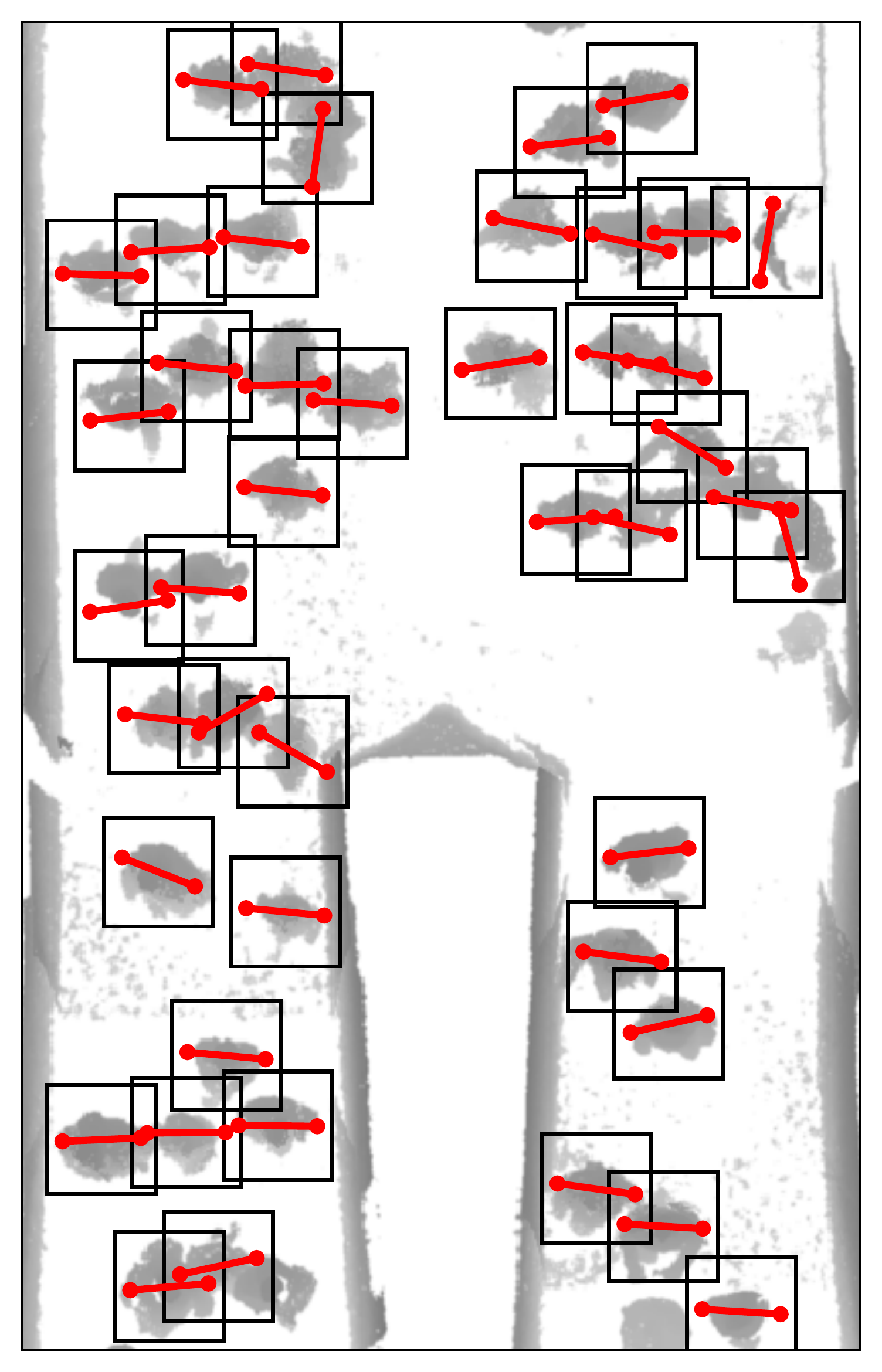}};
\begin{scope}[x={(img.north east)},y={(img.south west)}]
	\node [anchor = north west, text width=0.25cm, align=center, scale=1.2] at (-0.01, -0.01){(c)};
\end{scope}
\end{tikzpicture}

\caption{We measure and investigate the dynamics of shoulder
  orientation for walking pedestrians in real-life scenarios. Our
  measurements are based on raw data acquired via grids of overhead
  depth sensors, such as Microsoft \kinectTM~\cite{Kinect}. In (a,b)
  we report, respectively, a front and an aerial view of a data
  acquisition setup (similar to that in~\cite{corbetta2018large}). The
  sensors, of view the typical view cone is reported in (a), are
  represented in (b) as thick segments. In overhead depth images (c),
  the pixel value, here colorized in gray, represents the distance
  between each pixel and the camera plane: brighter shades are far
  from the sensor and, the darker the pixel color, the closer the
  pixel is to the sensor. Heads are, therefore, in darker shade than
  the floor. Through localization and tracking algorithms from
  ~\cite{corbetta2016fluctuations,PhysRevE.98.062310}, we extract
  imagelets centered on individual pedestrians (cf. imagelets
  annotated with ground truth in Fig.~\ref{fig:challenges}) for which
  we estimate orientations via the method introduced
  here.}\label{fig:sketch}
\end{figure}
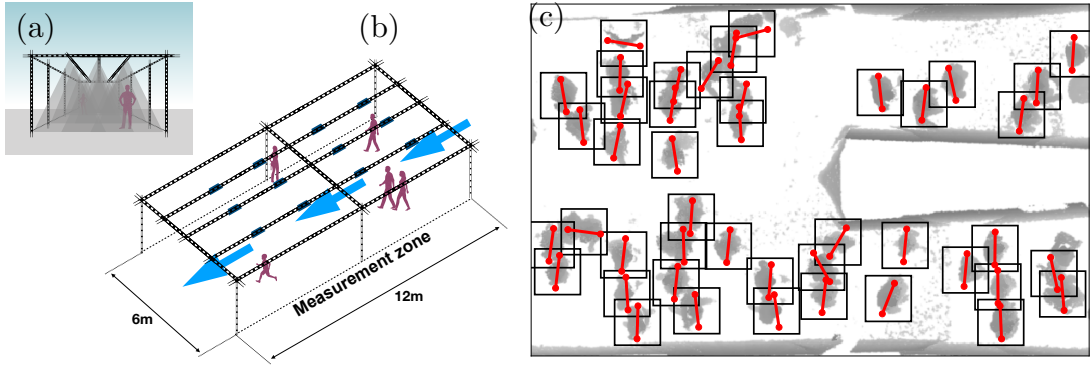

%% Explain why in real life.
The dynamics of shoulder-line rotation has been scarcely investigated
from a quantitative viewpoint. The data currently available is
extremely limited and has been acquired via few laboratory experiments
(e.g.~\cite{yamamoto2019body,Feliciani2016PedestriansRM}). Such
scarceness of accurate data hinders the capability of statistic
characterizations of the rotation dynamics beyond the estimation of
average properties, to include, e.g., fluctuations and rare events. We
believe that this is connected with the inherent technical complexity
of measuring body yawing accurately and in real-life
conditions. Real-life measurements campaigns, in fact, need to rely
only on non-intrusive imaging data (or alike) of pedestrians, and
cannot be supported by \textit{ad hoc} wearable sensors, such as
accelerometers~\cite{Feliciani2016PedestriansRM}. Indeed, even the
accurate estimation of the position of an individual in real-life, a
more ``macroscopic'' or ``coarser-scale'' degree of freedom than
orientation, is a recognized technical
challenge~\cite{DBLP:journals/ijon/BoltesS13}.  Since few years,
overhead depth-sensing~\cite{seer2014kinects,brscic2013person,
  corbetta2016fluctuations,PhysRevE.98.062310}, as used in this work,
has been successfully employed to perform accurate pedestrian
localization and prolonged tracking campaigns (see example in
Fig.~\ref{fig:sketch} and~\cite{corbetta2018large}). Overhead depth
data, not only allows privacy respectful data acquisition, but enables
also accurate position measurements even in high-density conditions
(for a highly-accurate algorithm leveraging on machine learning-based
analyses see, e.g.,~\cite{KronemanCorbettaPed18}).

\begin{figure*}[t!]
\centering
\snapshotrotation{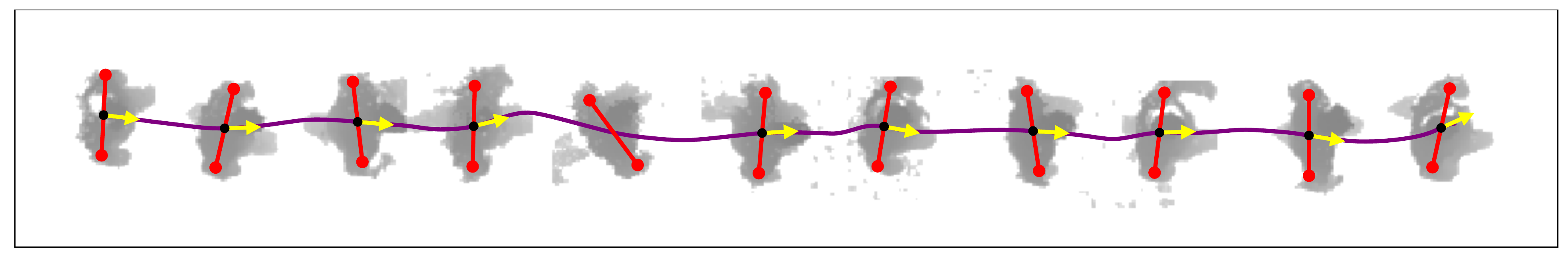}{15.67}{0.1}{a}{34}
\begin{adjustbox}{width=0.98\textwidth}
	\begin{tikzpicture}
		\node[anchor=north west,inner sep=0] (img) at (0,0) {\includegraphics{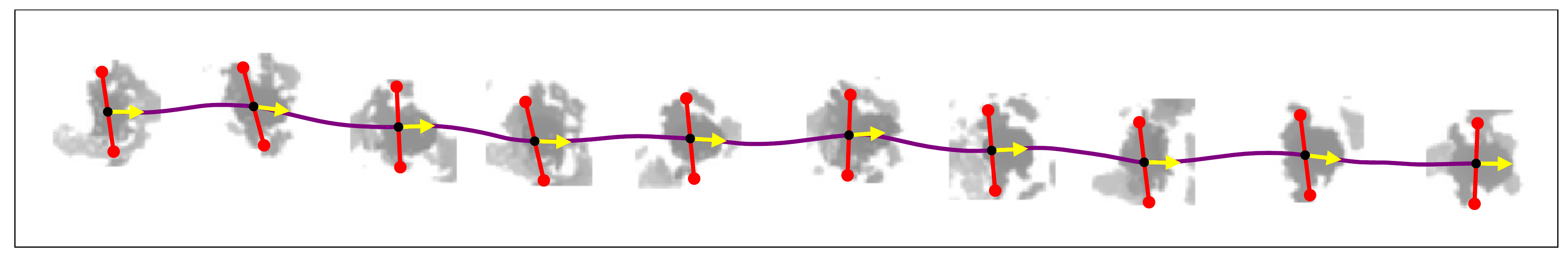}};
		\node [anchor = north west, scale=2.75] at (0.2, -0.2){(c)};
    	\end{tikzpicture}
\end{adjustbox}
\caption{(a,c) Pedestrian trajectories (purple) superimposed to depth
  snapshots (gray). Orientation estimates and local velocities
  (directions of motion) are reported, respectively, in red and
  yellow. We estimate shoulder orientation on a snapshot-by-snapshot
  basis, considering depth ``imagelets'' centered on a pedestrian. The
  sub-panel (b) reports an example of such an imagelet with the $x-y$
  coordinate system considered. We employ instantaneous direction of
  motion $\theta_v$ extracted from preexisting trajectory data as
  training labels for a neural network. This yields a reliable
  estimator for the orientation $\theta$, accurate even in cases
  challenging for humans, like in (c). Due to clothing, arms and body
  posture, presence of bag-packs or errors in depth reconstruction,
  the overhead pedestrian shape might appear substantially different
  from an ellipse elongated in the direction of the shoulders.
} \label{fig:challenges}
\end{figure*}

In this paper we propose a novel method to measure -in real-life
conditions and with very high accuracy- the shoulder rotation of
walking pedestrians. Our measurement method is based on a deep
Convolutional Neural Network~\cite{lecun2015deep} (CNN)
point-estimator which operates on overhead depth images centered on
individual pedestrians - from now on referred to as ``imagelets''.
Intuition suggests that pedestrians seen from an overhead perspective
have a well-defined elongated, elliptic-like, shape. In our
measurements this is true only in a small fraction of cases in which
pedestrians walk carrying their arms alongside the body. Conversely,
we found a majority of exceptions, impossible to address by hand-made
algorithms (cf. Fig.~\ref{fig:challenges}). This marks an ideal
use-case for supervised deep learning~\cite{lecun2015deep}.

\noindent It is well known that the high performance of Deep Neural
Network methods come also at the price of, often prohibitively,
labor-intensive manual annotations of training data (frequently in the
order of millions of individual images). Depending on the context, the
reliability of human annotations can furthermore be arguable, this is
the case whenever different experts are in frequent mutual
disagreement about the annotation value. Shoulder orientation in depth
imagelets falls in such a case. Here by relying on the strong
statistical correlation between individual velocity and body
orientation, we manage to produce potentially limitless annotations.
While walking on straight paths, our velocity direction is (on
average) in very good approximation orthogonal to our shoulder
line. On this basis, we can employ the velocity direction as a
singularly slightly imperfect, but correct on average, annotation for
the orientation. Notably, the zero-average residual error between the
velocity direction and the actual orientation gets averaged out as we
train our CNN point-estimator with gradient descent. This (self)
amends for annotation errors.

We investigate the orientation measurement accuracy of our method and
consider its error scaling vs. the size of the training set using
both real-life and synthetic depth imagelets. Combining extensive
training with the enforcement of $O(2)$ symmetry of the estimator, we
show that we can deliver an orientation estimator with an error as low
as $7.5$ degrees. Our tool enables us to characterize the stochastic
process that connects the instantaneous velocity direction to the
shoulder orientation. We show that the velocity orientation can be
well described by delaying the orientation dynamics through a
stochastic process centered on, about, $100\,$ms and with
Ornstein-Uhlenbeck (OU) statistics.

Conceptually speaking, although our tool has been devised for depth
imagelets, it can be easily extended to other computer vision-based
pedestrian tracking approaches and, more in general, can be used for
any system in which there is a statistical connection between
(average) individual ``particle'' velocity and (average) shape.

\begin{figure}[t!]
\centering
\begin{adjustbox}{width=0.2\textwidth}
	\begin{tikzpicture}
	\draw (0, 0) node[anchor=north west,inner sep=0] (img) {\includegraphics[width=0.10\textwidth]{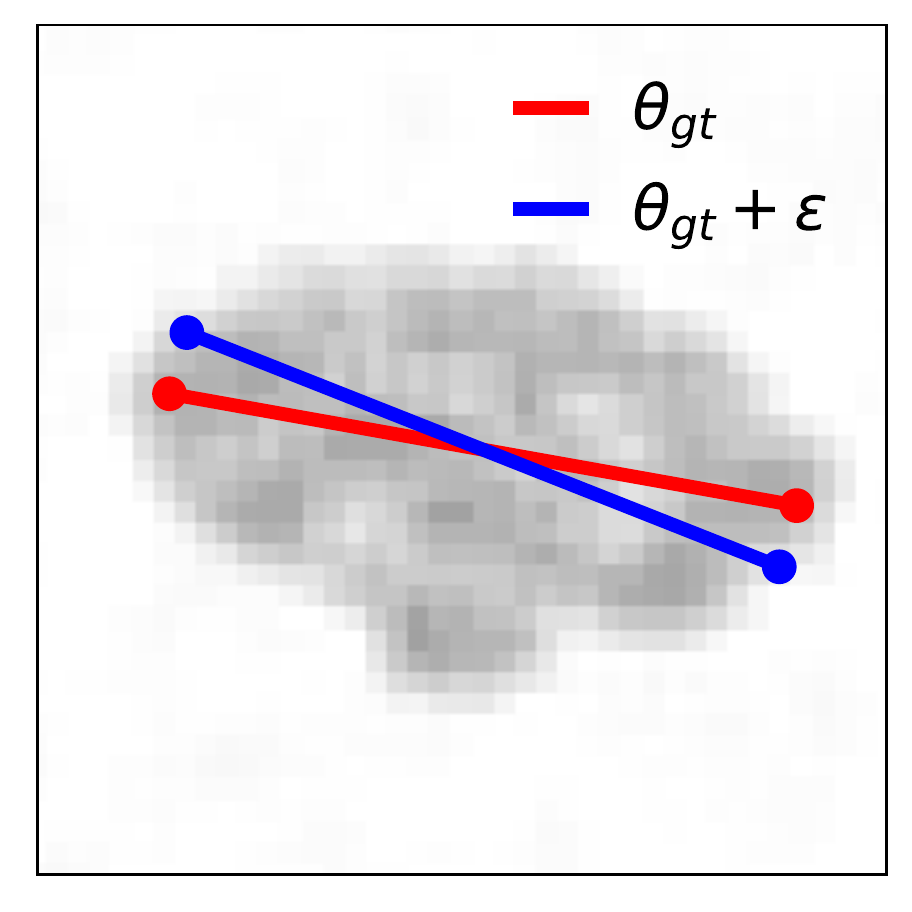}};
		\begin{scope}[x={(img.north east)},y={(img.south west)}]
			\node [anchor = north west, align=center, scale=0.65] at (.01, .01){(a)};
		\end{scope}
	\end{tikzpicture}%
\end{adjustbox}
\begin{adjustbox}{width=0.2\textwidth}
	\begin{tikzpicture}
	\draw (0, 0) node[anchor=north west,inner sep=0] (img) {\includegraphics[width=0.10\textwidth]{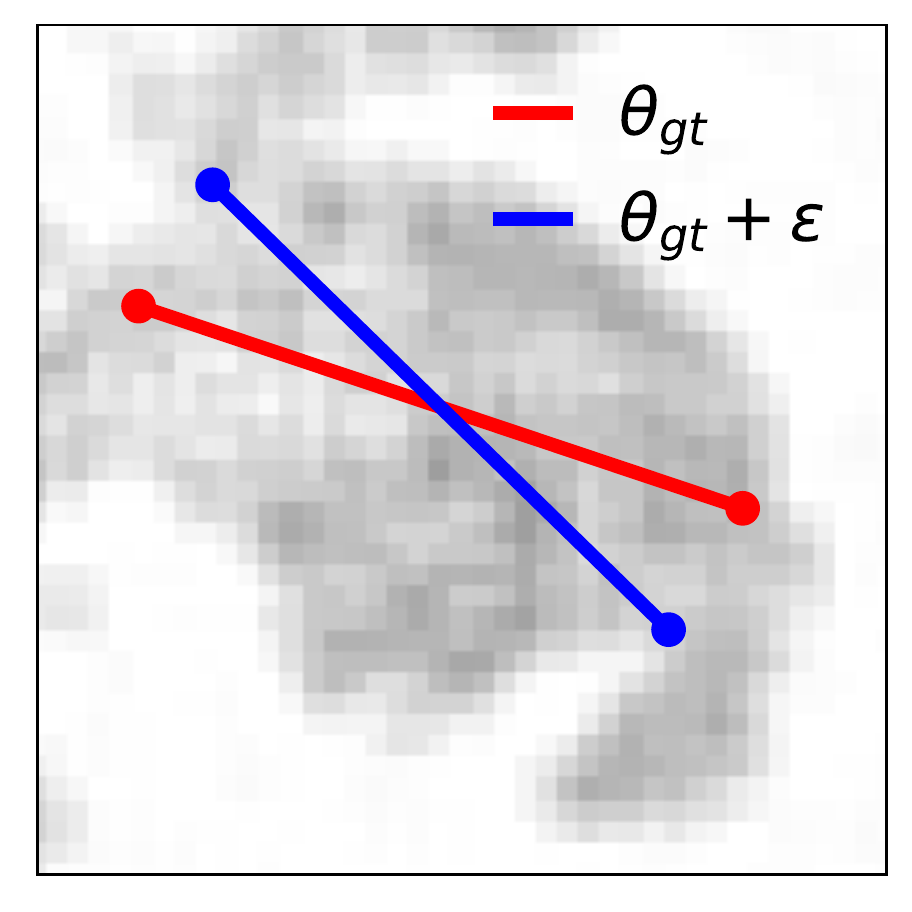}};
		\begin{scope}[x={(img.north east)},y={(img.south west)}]
			\node [anchor = north west, align=center, scale=0.65] at (.01, .01){(b)};
		\end{scope}
	\end{tikzpicture}%
\end{adjustbox}
\begin{adjustbox}{width=0.2\textwidth}
	\begin{tikzpicture}
	\draw (0, 0) node[anchor=north west,inner sep=0] (img) {\includegraphics[width=0.10\textwidth]{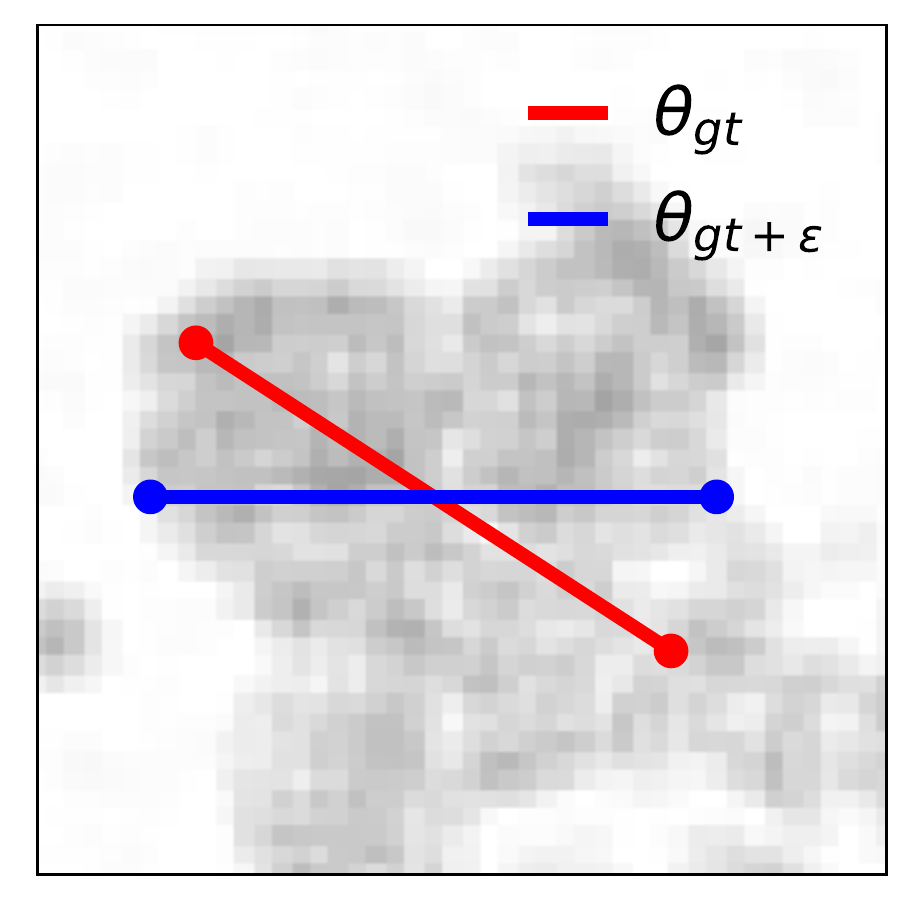}};
		\begin{scope}[x={(img.north east)},y={(img.south west)}]
			\node [anchor = north west, align=center, scale=0.65] at (.01, .01){(c)};
		\end{scope}
	\end{tikzpicture}%
\end{adjustbox}
\caption{Examples of synthetic imagelets that we employ to analyze the
  performance of our neural network. Contrarily to the real-life data,
  ground-truth orientation is available for synthetic data, enabling
  accurate validation of the estimations. The neural network is
  trained against labeled target data with predefined noise level
  ($\sigma = 20^o$) to imitate training with real-life imagelets and
  velocity target data.  Target data with predefined noise level and
  ground-truth orientation for validation are superimposed on the
  imagelets as blue and red bars
  respectively. 
} \label{fig:synthetic_examples}
\end{figure}
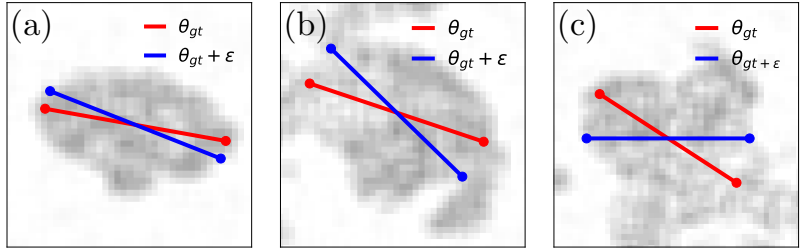

\section*{Orientation measurements: problem definition}\label{sect:or-meas}

Let $\Il$ be a overhead imagelet centered on a pedestrian, see
examples in Fig.~\ref{fig:challenges} (for convenience we
opt for imagelets of square shape, yet this is not a constraint).

We define the shoulder-line orientation angle, $\theta$, as the angle
between the direction normal to the shoulder-line and a fixed
reference, here the $y$ axis (direction $\vec{e}_y$, cf.
Fig.~\ref{fig:challenges}(a,b)). According to this definition, a
body rotation of $180^\circ$ leaves $\theta$ unchanged.  Thus, we aim
at a function $f$ such that
\begin{equation}
  \Il \xmapsto{f} \theta_o \approx \theta \in \Pj^1(\R) \cong [-\pi/2,\pi/2),
  \label{eq:estimator_definition}
\end{equation}
where $f(\Il) = \theta_o$ approximates the actual orientation $\theta$
(with $-\pi/2\equiv\pi/2$, i.e. $\theta$ is an element of the
real projective line $\Pj^1(\R)$, see e.g.~\cite{gowers2008princeton}).

We model the mapping $f$ via a deep neural network that we train in a
supervised, end-to-end, fashion (see structure in the Supporting
Information, SI). The network returns the estimate of $\theta$ as a
discrete probability distribution, $h_{\text{pred}}$, on $\Pj^1(\R)$
(quantized in $B=45$ uniform bins, $4^\circ$ wide, via soft-max
activation function in the final layer). We retain the
$\Pj^1(\R)$-average (``circular average'') of the distribution
$h_{\text{pred}}$, as final output. It formulas, the output $\theta_o$
reads
\begin{equation}
  \theta_o = \E_{\theta'\sim h_{\text{pred}}, \Pj^1(\R)}[\theta'];
  \label{eq:Proj-avg}
\end{equation}
we leave the details to the SI.

We train with orientation data with a ``two-hot'' encoding: each
orientation $\theta$ is unambiguously represented in terms of a
probability distribution non-vanishing on (up to) two adjacent angular
bins (we chose ``two-hot'' in opposition to the typical one-hot
training data for classification problems, in which the annotations
are Dirac probability distributions on the ground-truth class). We
will refer to this encoding, that avoids quantization errors, as
$h_2(\theta)$ (we observed no strong sensitivity on the number of bins
when these were more or equal than $10$). As usual, we
use a cross-entropy loss, $\loss(\cdot,\cdot)$.\\

We employ pedestrian velocity information to tackle the need for huge
amounts of accurately annotated data to train the free parameters
of the deep neural network (usually in the millions, $\approx 1.3M$ in
our case).

Let $\theta_v(t) \in \Pj^1(\R)$ be the angle between the walking
velocity and a reference at time $t>0$, i.e.
\begin{equation}
  \theta_v(t) = \angle(\vec{v}(t),\vec{e}_y),
  \label{eq:theta-v}
\end{equation}
where $\vec{v}(t)$ is the instantaneous velocity, and
$\angle(\cdot,\cdot)$ denotes the angle comprised the directions in
its argument (with $\pi$-periodicity).  Our shoulder line is
most-frequently, and in very good approximation, orthogonal with
respect to the walking velocity, i.e.
\begin{equation}\label{eq:vel-approx-orient}
  \theta\approx\theta_v.
\end{equation}
Therefore, velocities provide a meaningful ``proxy'' annotation for
orientation. We used the ``approximately equal'' sign
in~\eqref{eq:vel-approx-orient} because we can have frequent, yet
small, disagreements between velocity and orientation. These can be
due to small loss of alignment between the two (e.g. because something
attracted our attention) or they can be due to inaccuracies, e.g., in
the velocity measurements. It is also possible, yet less likely, that
velocity and orientation remain misaligned for longer time
intervals. This holds, e.g., for people walking sideways. We retain
these as rare occasions, which we expect to occur symmetrically for
both left and right sides, with no relevant weight in our training
dataset. This hypothesis reasonably holds on unidirectional pedestrian
flows happening on rectilinear corridors, but might be invalid in
case, e.g., of curved paths.
% In the latter conditions, lanes can form and pedestrians might sistematically avoid one another by rotating the shoulders in the same direction.
Formally, for a walking person, we model the relation
in~\eqref{eq:vel-approx-orient} as
\begin{equation}\label{eq:velocity-proxy}
    \theta = \theta_v + \epsilon,
\end{equation}
with $\epsilon$ being a small, symmetric, and zero-centered residual.

We train our neural network using the labels $h_2(\theta_v)$ as a
proxy for $h_2(\theta)$. The training process aims at the minimization
of the (average) loss
$\E_{\theta_v}[\loss(h_{pred},h_2(\theta_v))]$. As such, the output
$h_{pred}$ converges to the distribution of annotations of similar
imagelets, whose average is the correct point-estimation of the
orientation:
\begin{equation}
\theta_o \approx \E_{\theta' \sim \theta_v + \epsilon, \Pj^1(\R)}[\theta'] = \E[\theta - \epsilon] = \theta.
\end{equation}
We refer to the SI for a formal proof with simplifying assumptions and
a simulation-based proof in the general case.

Finally, once a pedestrian with shoulder orientation $\theta$ rigidly
rotates around the vertical axis by an angle $\alpha$, their
orientation becomes $\theta + \alpha$. Similarly, ``mirroring'' a
pedestrian around the $\vec{e}_y$ direction, their
orientation changes sign.
The map $f$ must respect such symmetry with respect to the group
orthogonal transformations, $O(2)$ (see, e.g.,~\cite{grove39classical}), i.e., in formulas it must hold
\begin{equation}
  f(\phi \Il) = (f(\Il) + \alpha)\det(\phi)
  \label{eq:f-sym}
\end{equation}
for all transformations $\phi=\Phi R_\alpha \in O(2)$, that
concatenate a rotation of $\alpha$, $R_\alpha$, and, possibly, a
reflection (i.e. $\Phi \in \{\text{Id}, J\}$, respectively the
identity and the reflection, from which the sign change given by the
determinant of the transformation: $\det(\phi)=\det(\Phi)=\pm 1$).

Symmetries in neural networks are often injected at training time, by
augmenting the training set by all the symmetry group orbits.
Similarly, we include multiple copies of the same imagelets with
multiple random rotations with and without flipping. This also ensures
that the training set spans $\Pj^1(\R)$ uniformly. Yet, this does not
yield a strictly $O(2)$-symmetric estimator (\eqref{eq:f-sym}). We
further enforce this symmetry by constructing a new map, $\tilde f$,
as the $O(2)$-group average of $f$, which is thus strictly
respecting \eqref{eq:f-sym}. In formulas it holds
\begin{align}
  \label{eq:O2avg}
  \tilde{f}(\Il) &= \frac{1}{|O(2)|}\int_{O(2)}(f(\phi \Il) - \alpha)\det(\phi)\, d\phi\\
  &=\frac{1}{2\pi}\int_{0}^{2\pi}\frac{f(R_\alpha\Il) - f(JR_\alpha \Il)}{2} - \alpha\, d\alpha,\label{eq:Otwo-avg} 
\end{align}
we leave the proof of this identity, the $O(2)$-symmetry of $\tilde f$
and further details on $\Pj^1(\R)$-averages to the
SI. In the following, we consider approximations of the integral in
\eqref{eq:Otwo-avg} by equi-spaced and random sampling of $O(2)$.

\begin{figure*}[h!]
\centering
\figmerge{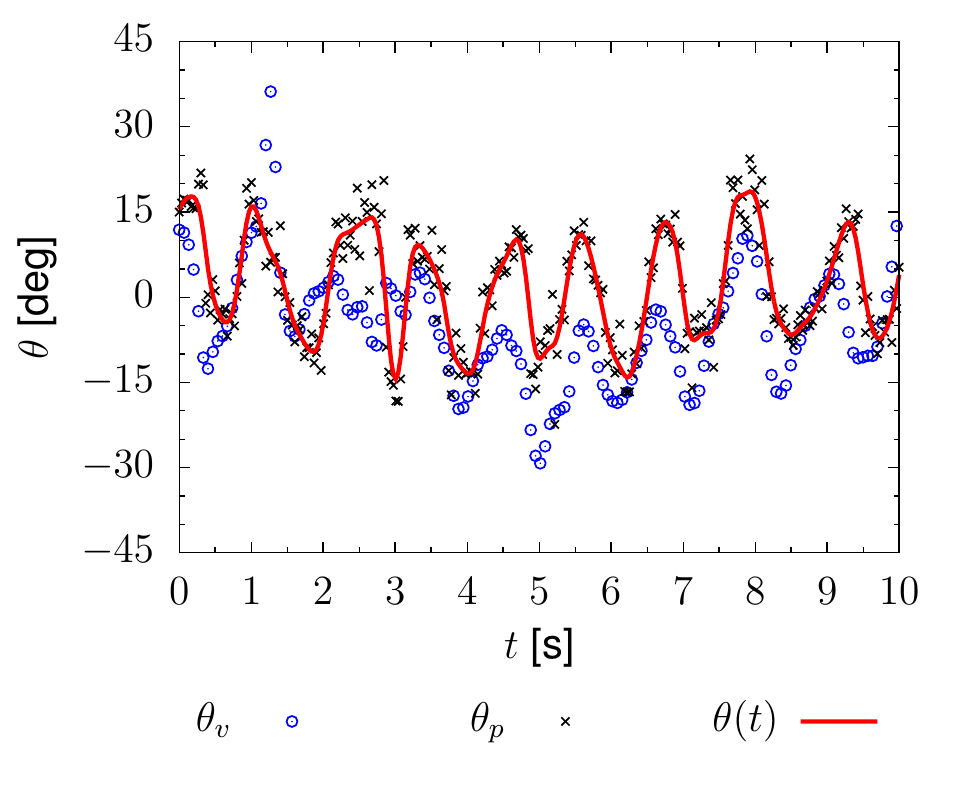}{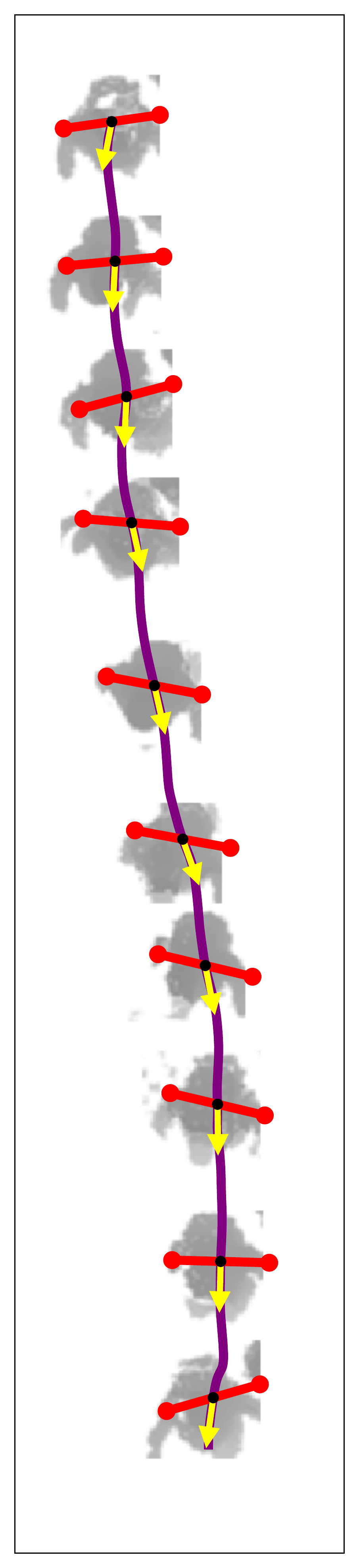}{a}{3.0cm}{$x$}{$y$}{0.045}
\figmerge{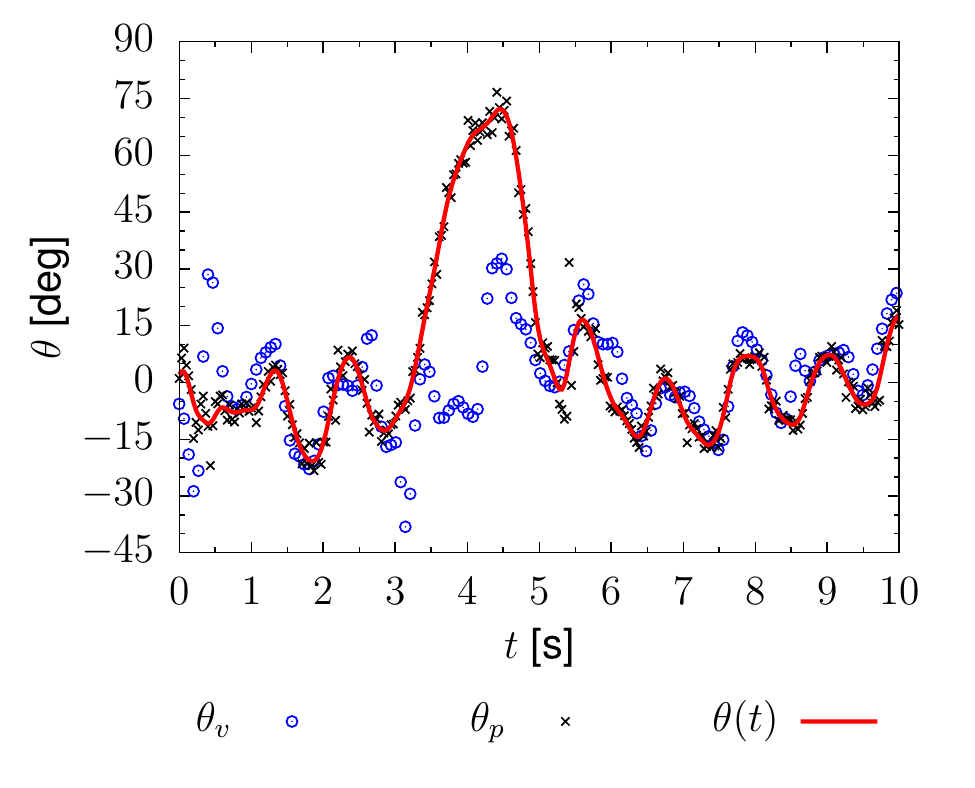}{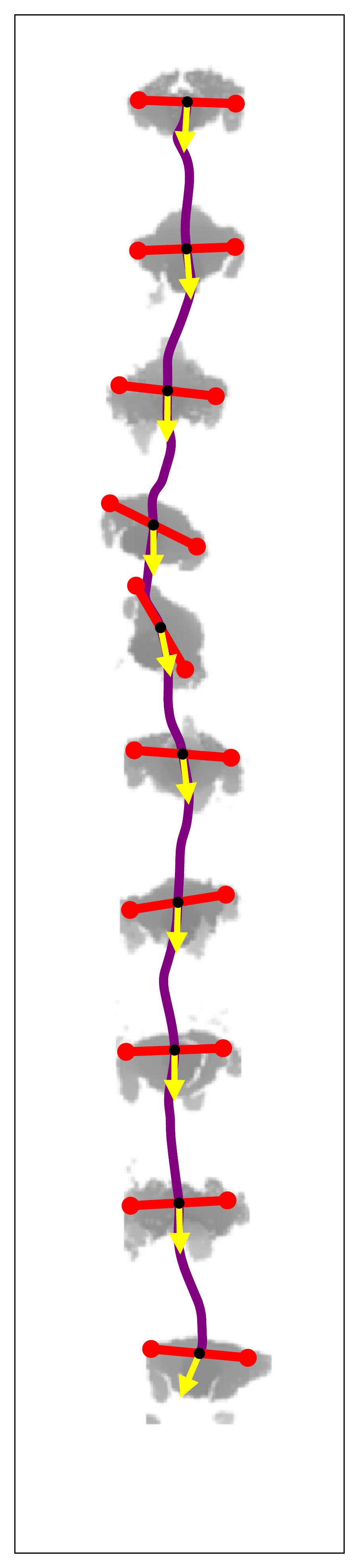}{b}{3.0cm}{$x$}{$y$}{0.045}
\figmerge{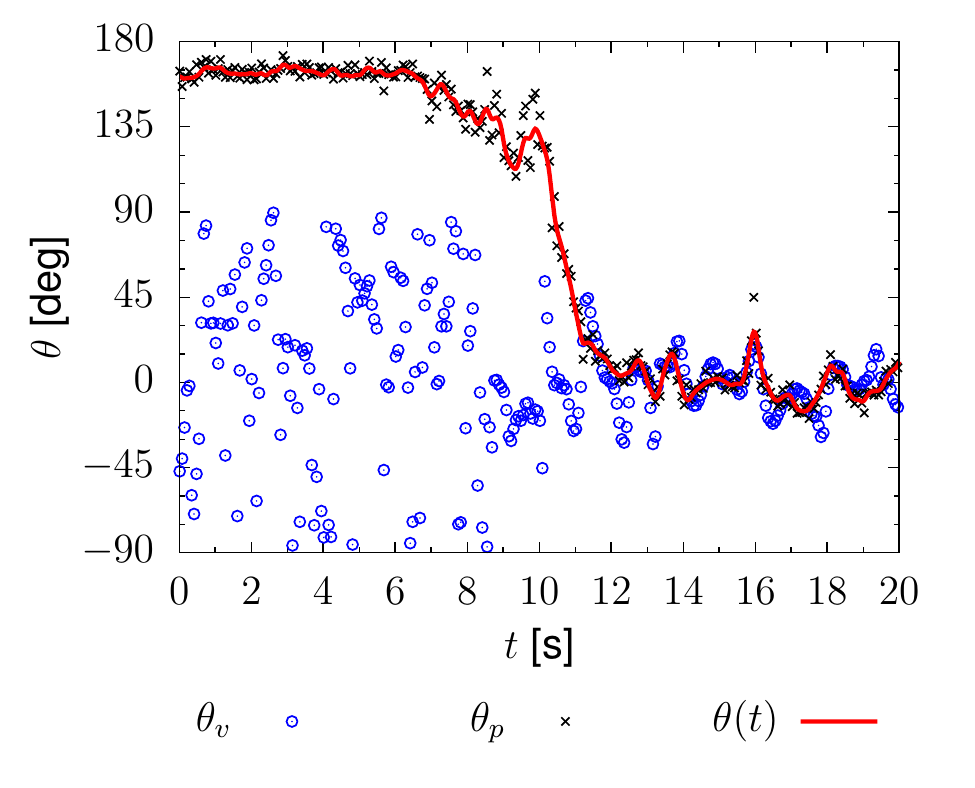}{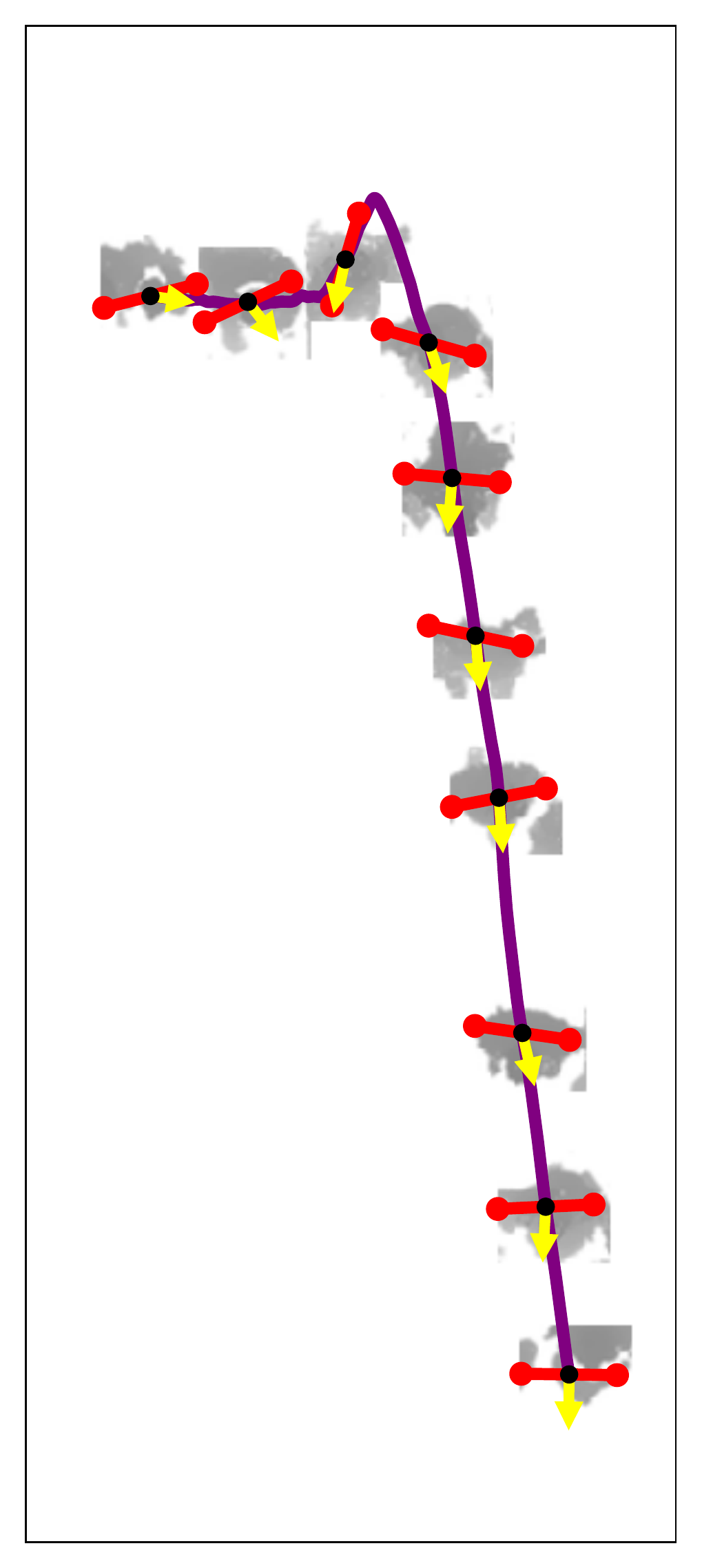}{c}{3.05cm}{$t$}{$y$}{0.038}
\figandor{\includegraphics[width=0.3\textwidth]{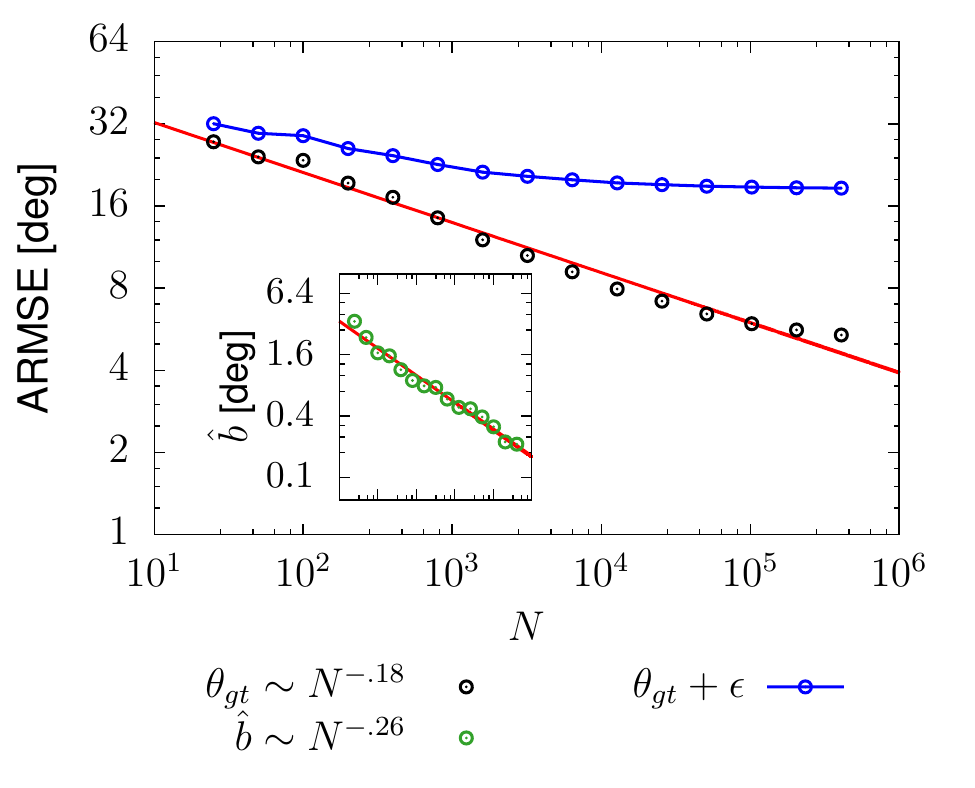}}{d}
\figandor{\includegraphics[width=0.3\textwidth]{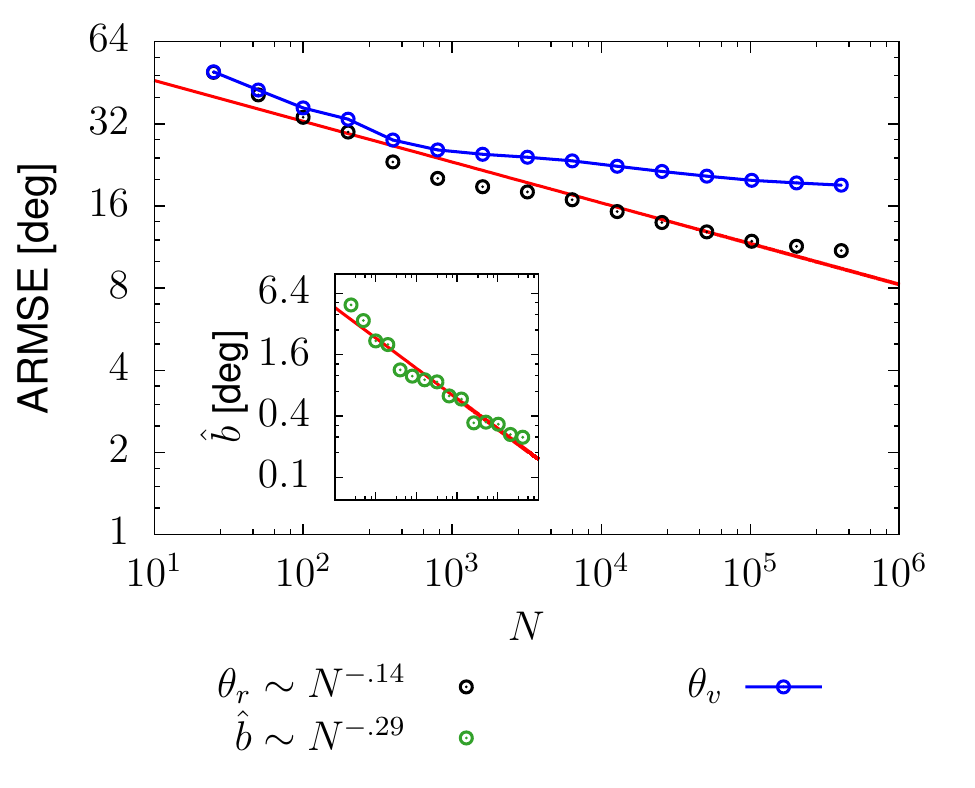}}{e}
\figandor{\includegraphics[width=0.3\textwidth]{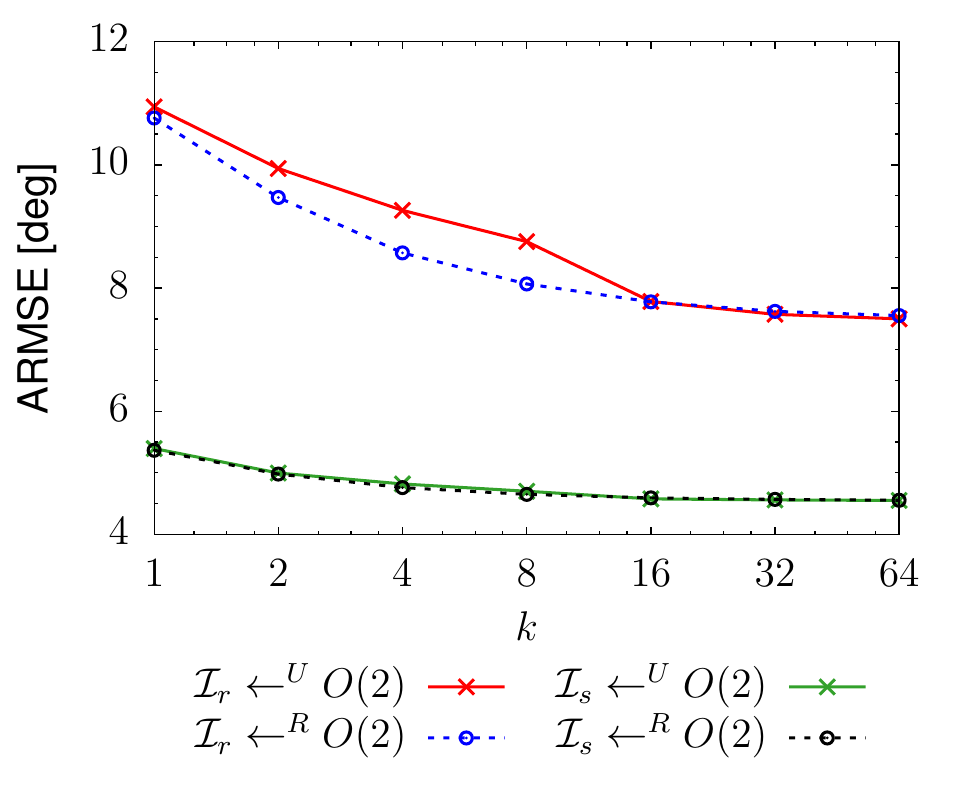}}{f}
\caption{(a-c) Velocity direction and shoulder orientation signal, for
  three trajectories collected in real-life (depth maps sequences
  similar to Fig.~\ref{fig:challenges} are on the right of each
  panel). We report the instantaneous values of velocity (obtained
  from tracking) and orientation, $\theta_v$ and $\theta_p$, and the
  continuous orientation signal $\theta(t)$ (low-pass filter of
  $\theta_p$). Orientation has been computed via our CNN trained on
  $30$ hours of real-life velocity data. Panel (a) reports a typical
  pedestrian behavior, where $\theta_v(t)$ and $\theta(t)$ oscillate
  ``in sync'' (frequency $f \approx$ $0.8\,$Hz) following the
  stepping. Our tool resolves correctly also rare sidestepping events
  or orientation of standing individuals in which the signals are out
  of sync.  Panel (b) shows a pedestrian rotating their body, possibly
  observing their surroundings, while maintaining the walking
  direction.  Panel (c) shows an individual initially standing, then
  performing a $150^o$ body rotation and finally walking away. In this
  case, the velocity $\theta_v$ is undefined for time $t<10\,$s as
  there is no position variation (so the high noise in $\theta_v$;
  cf. $yt$-diagram in which the spatial coordinates are constant in
  the horizontal segment of the trajectory).  (d-f) Prediction
  performance of the network, $f$ (~\eqref{eq:estimator_definition}),
  in case of artificial imagelets (d) and real-life data
  (e). We train with datasets of increasing size ($N$,
  $x$-axis). We report the Root Mean Square Error of the predictions
  averaged over $M=32$ independent training of the networks (ARMSE,
  \eqref{eq:pred-ARMSE}) and, in the inset, the average bias, $\hat b$
  (\eqref{eq:pred-bias}).
  The test sets used to compute the indicators include, for (d), $25$k
  unseen synthetic images with error-free annotation ($\theta_{gt}$)
  and, for (e), $25$k unseen real-life imagelets, annotated considering
  low-pass filtered high-resolution orientation estimates, $\theta_r$,
  obtained with our neural network trained with $1\,$M samples and
  $O(2)$-group averaging. The
  bias, in both cases, decreases rapidly below $0.1^\circ$. The ARMSE,
  for the networks trained with the largest dataset approaches,
  respectively, $5.5^o$ and $11^o$, as $N$ grows. For both ARMSE and
  bias, we report the fitted exponents characterizing the error
  converge in the label.
  We complement the evaluation of the ARMSE considering noisy labels
  ($\theta_{gt} +\epsilon$ for case (d) and $\theta_v$ for case
  (e)). In case (d) the ARMSE saturates consistently with the level of
  noise in the labels (cf. SI). In case (e), the ARMSE approaches a
  saturation point at about $20^o$. This reflects the random
  disagreement between velocity and orientation. (f) Performance can
  be further increased by enforcing $O(2)$-symmetry of the orientation
  estimator, map $\tilde f$, \eqref{eq:O2avg}. In panel (f) we
  consider maps $\tilde f$ built from the networks trained with the
  largest training datasets from (d,e) ($N\approx 0.5\,$M), both for
  the synthetic and real-life cases, vs. the number of samples used
  for the group average, $k$. We consider both uniform and random
  sampling of $O(2)$ (superscript $U$ and $R$ respectively). The group
  average further reduces the ARMSE from $5.5^o$ to $4.5^o$ in case of
  synthetic imagelets $\mathcal{I}_s$ (no observable difference
  between uniform and random sampling), and from $11^o$ to $7.5^o$ in
  case of real-life imagelets $\mathcal{I}_r$, with higher performance
  in case of random sampling for $k < 16$.
}\label{fig:examples-of-tracks}\label{fig:sigma_of_xbar}
\end{figure*}

\section*{CNN: training and testing}\label{sec:NN-training}

We consider two types of training/testing imagelets: algorithmically
generated, ``synthetic'', imagelets, of which the orientation angle
$\theta$ is known, and real-life imagelets.  In the first case we
mimic a velocity-based training by adding a centered noise to labels
known exactly (following~\eqref{eq:velocity-proxy}). In the second
case, as we have no manually annotated ground truth, of which the
accuracy would nevertheless be debatable, we propose a validation
based on the convergence towards low-pass filtered orientation
signals. In both cases, we show that the average prediction error (ARMSE,
\eqref{eq:pred-ARMSE}) is about $7.5^\circ$ degrees or, possibly,
lower, should the training set size $N$ be large enough. Specifically,
the datasets are as follows:

\noindent \textit{Synthetic dataset.} We generate synthetic imagelets
  mimicking the overhead shape of people in terms of a superposition
  of two ellipses: one for the body/shoulder, $E_b$, and another one,
  $E_h$, at lower depth values (i.e. higher on the ground), for the
  head. We report examples of such imagelets in
  Fig.~\ref{fig:synthetic_examples}, while the details of the
  generation algorithm are left to the SI.

  \noindent By construction, the rotation angle $\alpha_{b}$ of $E_b$
  represents the pedestrian orientation, i.e. it is the ground truth
  for the training. We train the network with such synthetic imagelets
  and a small centered Gaussian noise
  $\epsilon \sim \mathcal{N}\left(0^o, 18^o\right)$ superimposed to
  $\alpha_{b} = \theta_{gt}$ to imitate velocity-based
  training. Hence, we train using labels $\alpha_{gt} + \epsilon$
  while we validate with $\alpha_{gt}$ (cf.
  Eq.~\ref{eq:velocity-proxy}).\\

\noindent \textit{Real-life dataset.} We consider depth images and
  velocity data from a real-life measurement campaign conducted 
  during a city-wide festival (GLOW) in Eindhoven, The Netherlands, in
  Nov. 2017. The measurements involve a uni-directional crowd flow
  passing through a corridor-shaped exhibit (tracking area:
  $12m \times 6m$), for further details see~\cite{corbetta2018large}.
  The dataset leverages on high-resolution individual localization and
  tracking based on overhead depth images (as in
  Fig.~\ref{fig:sketch}) and with $30\,$Hz time sampling. The
  localization and tracking algorithms employed are analogous to what
  employed in previous
  works~\cite{corbetta2016fluctuations,PhysRevE.98.062310}. To ensure
  that our velocity data provides a well-defined proxy for
  orientation, we restrict to pedestrians having average velocity
  above $0.65\,$m/s. Moreover, for each trajectory we extract
  imagelets and velocity data with a time sampling of
  $\Delta T \approx 0.5\,$s, which increases the independence between
  training data.
  Additionally, we apply random rotations % (with
  % bicubic interpolation)
  and random horizontal flips to all imagelets (and, correspondingly,
  to labels). This aims at training with a dataset uniformly
  distributed on $\Pj(\R^1)$. 

  \noindent In absence of ground truth, we build our test set as
  follows: we rely on our neural network trained with $1\,$M different
  imagelets (i.e. twice as much the largest training dataset
  considered in Fig.~\ref{fig:sigma_of_xbar}(d,e), on which we perform
  random augmentation and final $O(2)$-averaging of the operator),
  hence the most accurate, to make orientation predictions over
  complete pedestrian trajectories. As an orientation signal
  $\theta(t)$ needs to be continuous in time, we smoothen the
  predicted $\theta(t)$ in time (low-pass Butterworth filter~
  \cite{butter} of order $n=1$, cutoff frequency $c_f=2.0\,$Hz and
  window length $l=52$) to
  eliminate random noise.\\

We assess the prediction performance as the training set size, $N$,
increases. To compute exhaustive performance statistics, for every
$N$, we train the network on $M$ independent datasets.
Given a reference orientation (e.g. ground truth),
$\theta_r$, for imagelet $\Il$ from dataset $D_k$ ($k=1,2,\dots,M$),
we consider two statistical indicators: (S1) the average prediction
bias, $\hat{b}$, evaluated as the root-mean-square (among the $M$
networks) of the average
\begin{equation}\label{eq:pred-bias}
  \hat{b} = \sqrt{\frac{1}{M}\sum_{k=1}^M{ \left(\E_{\Il \in D_k}[\theta_o - \theta_r]\right)^2 }};
\end{equation}
(S2) the average root-mean-square error (ARMSE), i.e. the average (on
  the $M$ networks) of the individual RMS error
\begin{equation}\label{eq:pred-ARMSE}
  \text{ARMSE} = \frac{1}{M} \sum_{k=1}^M \left(\E_{\Il \in D_k}[ (\theta_o - \theta_r)^2  ]\right)^{\frac{1}{2}}.
\end{equation}

\section*{Results}
In Fig.~\ref{fig:examples-of-tracks}(a-c), we report the orientation
signals as estimated by the networks in three different real-life
contexts. The network is capable of accurate predictions that, as
expected, are independent of the actual instantaneous velocity. Hence,
it remains accurate in case of a pedestrian walking sideways
(Fig.~\ref{fig:examples-of-tracks}(b)), in which the orientation
signal loses temporarily coupling with the velocity orientation and in
case of a pedestrian temporarily stopping and standing
(Fig.~\ref{fig:examples-of-tracks}(c)), in which the velocity
orientation is even undefined (note that these cases were excluded
from the training).

We include in Fig.~\ref{fig:sigma_of_xbar}(d-f) the values of average
prediction bias and ARMSE as the training set size increases, in case
of synthetic and real-life imagelets (respectively, in panels (d) and
(e)). In both cases the network performance increases with $N$, with
slightly slower convergence rate for the ARMSE for the real-life
dataset, which is likely more challenging to learn than the synthetic
one. In both cases the predictions are free of bias
(cf. sub-panels). With the largest number of training imagelets
considered ($N\approx 5\cdot 10^5$), we measured an ARMSE of about
$5^\circ$ for the synthetic data and $11^\circ$ for the real-life
data. We managed to further reduce this error to, respectively,
$4.5^\circ$ and $7.5^\circ$ by enforcing $O(2)$ symmetry. Note that we
could trivially apply \eqref{eq:O2avg} as we are in a bias-free
context, else a systematic correction for the bias would have been
necessary. In Fig.~\ref{fig:sigma_of_xbar}(f), we report the network
performance as we approximate better and better the $O(2)$ group
average.

\begin{figure}[t!]
\centering
\includegraphics[width=0.4\textwidth]{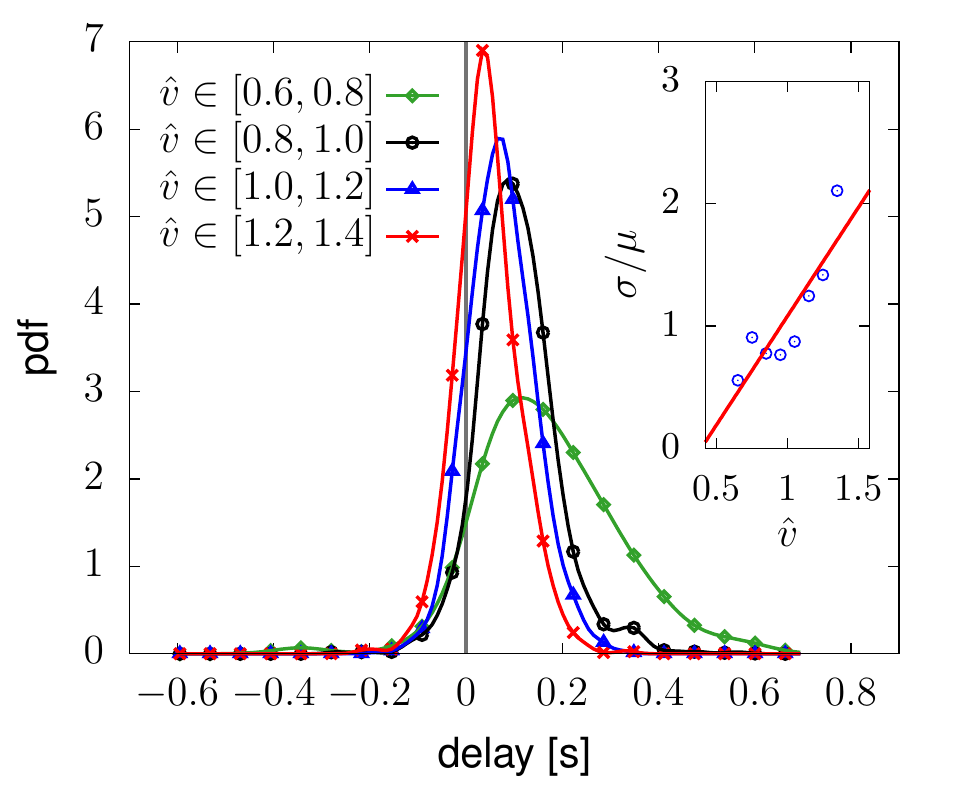}
\caption{Probability distribution function of the delay time between
  the shoulder orientation, $\theta(t)$, and velocity orientation,
  $\theta_v(x)$, signals for different average velocities, $\hat
  v$. As the average velocity grows, the average delay and the delay
  fluctuations reduce. The inset reports the ratio between the
  standard deviation, $\sigma$, and the average, $\mu$, of the delay
  as a function of $\hat v$ (Measurements from $78$k trajectories
  (all not exceeding a maximum orientation of
  $\pm60^\circ$, $20\,$ hrs of data), acquired during the GLOW
  event.~\cite{corbetta2018large}).\label{fig:delay-vs-velocity}}
\end{figure}

\begin{figure*}[t!]
\centering
\figandor{\includegraphics[width=0.3\textwidth]{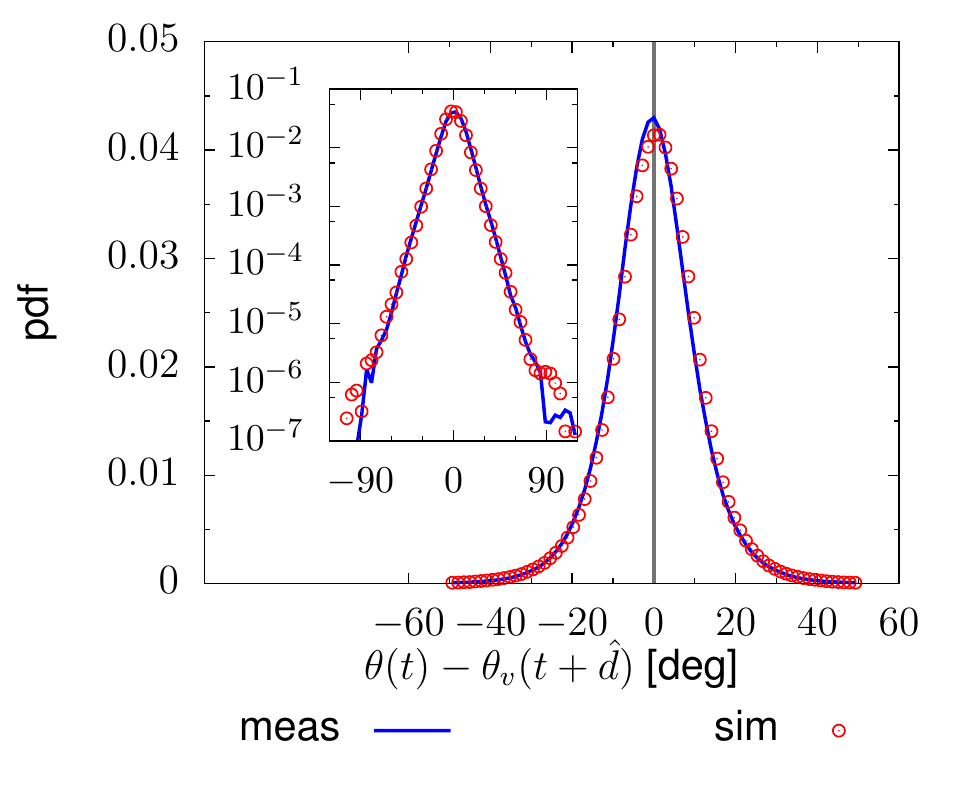}}{a}
\figandor{\includegraphics[width=0.3\textwidth]{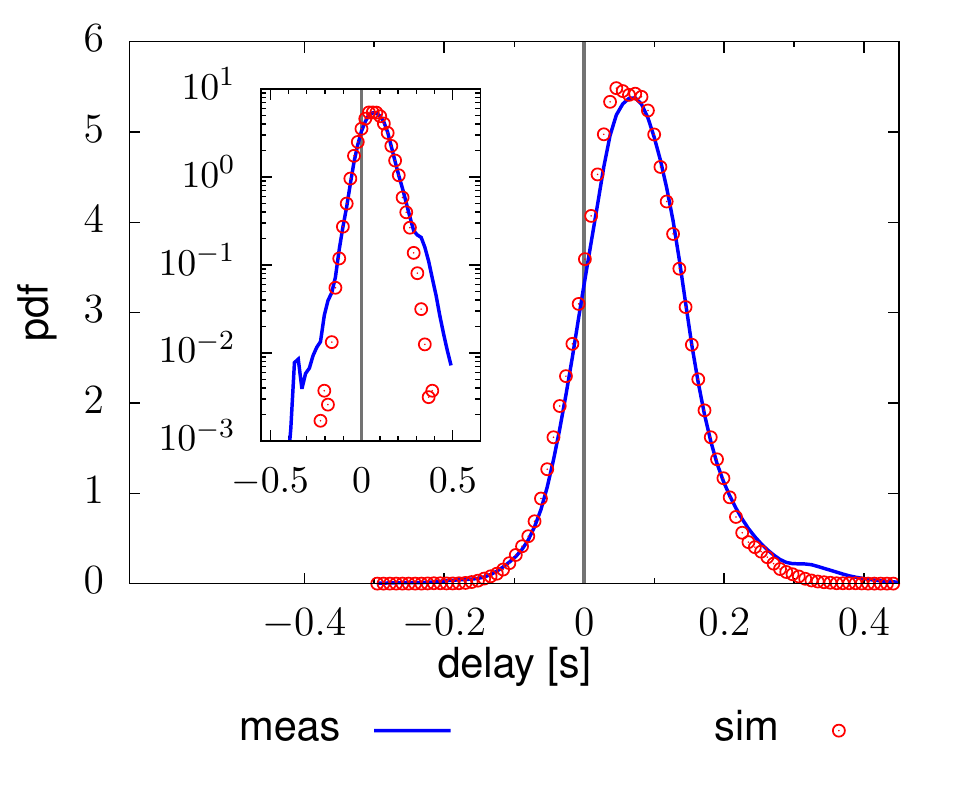}}{b}
\figandor{\includegraphics[width=0.3\textwidth]{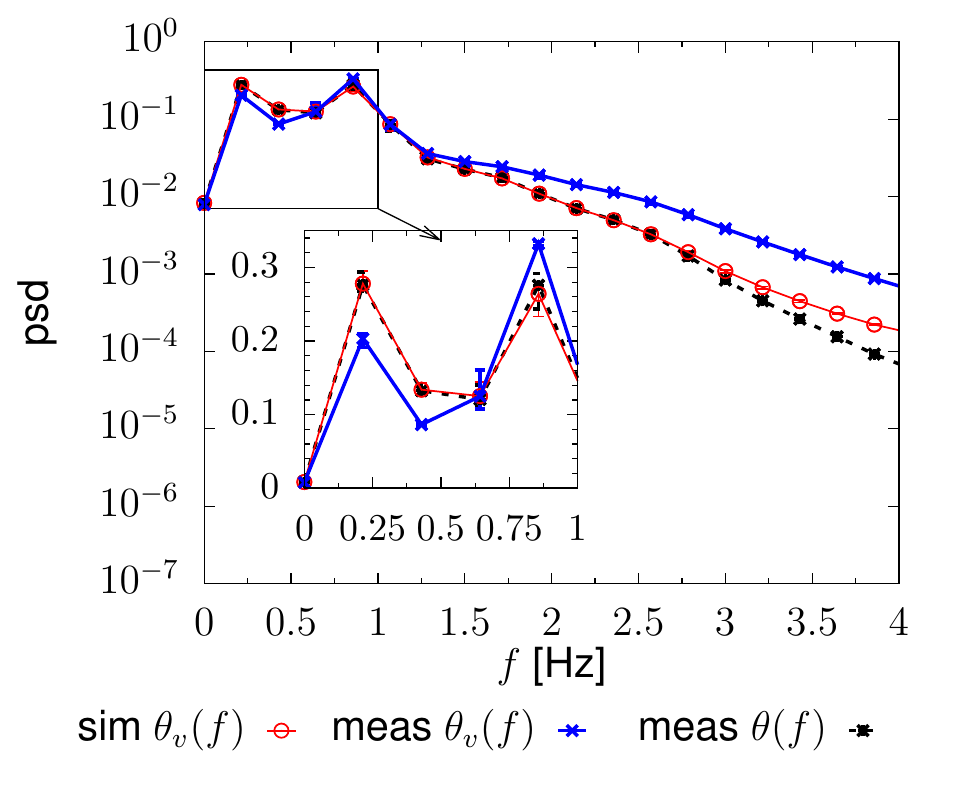}}{c}
\caption{Comparison between simulations (red dots) and real-life
  measurements (blue dotted line). We build velocity direction signals
  $\theta_v(t)$ on top of delayed orientation measurements
  $\theta(t-d(t))$, where the delay $d(t)$ is modelled by a OU random
  process (cf. \eqref{eq:theta-model},
  \eqref{eq:OU-process}). Measurements have been acquired during the
  GLOW event ($36$k trajectories, restricting to people keeping
  normal average velocity $\hat v \approx 1.3\,$m/s). In (a) we report
  the probability distribution function (pdf) of the difference
  between velocity direction and orientation shifted in time by the
  average delay, $\hat{d} = 0.1\,$s. In (b), analogously to
  Fig.~\ref{fig:delay-vs-velocity}, we report a PDF of the delay time
  between $\theta(t)$ and $\theta_v(t)$. The insets in (a,b) report
  the data in semi-logarithmic scale. For both these quantities we
  observe excellent agreement among simulations and measurements for
  (a,b). Panel (c) shows the velocity direction signals' grand average
  power spectral density (psd) of $\theta_v(t)$ and $\theta(t)$.
  Our model modifies the psd only at high frequencies. As an effect,
  the most energetic components of the velocity orientation, around
  $0.2\,$Hz and $1\,$Hz, remain, respectively slightly under and
  slightly over-represented.  }\label{fig:simulations}
\end{figure*}

\section*{Real-life orientation dynamics}
\label{sect:real-life-dynamics}

We are now capable of investigating with high-resolution, and in
real-life conditions, the connection between shoulder orientation and
velocity direction - which, in the previous sections, we reduced to
the error term $\epsilon$. In particular, we can characterize a
stochastic delay signal, $d(t)$, which allows us to model the relation
between velocity and orientation as
\begin{equation}
  \theta_v(t) = A\, \theta(t-d(t)),
  \label{eq:theta-model}
\end{equation}
where $A$ is a positive constant.

First, thanks to the high-accuracy of our tool, we measure a
velocity-dependent delay between velocity orientation and shoulder
orientation whose probability distribution function is in
Fig.~\ref{fig:delay-vs-velocity} (see SI for details on the delay
measurement algorithm). The velocity orientation follows in time the
shoulder yawing, with a delay that decreases (on average) between
$160\,$ms and $100\,$ms as the average walking velocity, $\bar v$,
increases from $0.6\,$m/s to $1.4\,$m/s (respectively walking speed
values in leisure and normal walking regimes, see,
e.g.~\cite{fruin1987BOOK}).

The structure of $d(t)$ appears well-modeled by a OU process:
\begin{equation}
  \dot{d}(t) = -\frac{\hat d - d(t)}{\tau} + \xi \dot{W},
  \label{eq:OU-process}
\end{equation}
where $\hat d>0$ is the average delay
($\hat d \approx \langle d(t) \rangle$), $\tau>0$ is the OU time-scale
and $\xi>0$ is the intensity of the $\delta$-correlated white noise
$\dot{W}$. In particular, in Fig.~\ref{fig:simulations} we compare
statistical observables of measurements and simulations considering
the case of normal walking speed (average velocity
$\hat v \approx 1.3\,$m/s), of which we retain the measured
orientation signals, $\theta(t)$, as a basis for
\eqref{eq:theta-model} (simulation parameters: $A=1.85$,
$\hat d =0.08\,$s, $\tau=1.2\,$s and $\xi=1.85$). In
Fig.~\ref{fig:simulations}(a), we report the pdf of the difference
between orientation and velocity orientation when one is shifted in
time by, $\hat d$, to compensate for the average delay. Measurements
and simulations, in excellent mutual agreement, follow a Gaussian
statistics. Thanks to a stochastic delay, we achieve a very good
quantitative agreement in the delay distributions
(Fig.~\ref{fig:simulations}(b)). In Fig.~\ref{fig:simulations}(c), we
report the Power Spectral Density (psd) of $\theta_v(t)$ and
$\theta(t)$ computed by averaging all the psds obtained from
individual velocity direction and orientation signals. We observe that
the stochastic delay does not substantially modify the psd of
orientations, especially at low frequencies.  As an effect, the peak
around $f=1\,$Hz, connected with the walking fluctuations remains
slightly underestimated, while larger scale fluctuations
($f\approx 0.1\,$Hz), effectively not modeled by
\eqref{eq:OU-process}, are over-represented.

\section*{Discussion}
In this paper we presented an extremely accurate estimator for the
pedestrian shoulder-line orientation based on deep convolutional
neural networks. We leveraged on statistic aspects of pedestrian
dynamics to overcome two outstanding issues related to deep networks
training: the labor-intensive annotation of training data in
sufficient amounts (generally millions of images) and the accuracy
of annotations in non-trivial contexts.

Thanks to the strong statistical correlation of shoulder-line and
velocity direction, which are typically orthogonal, we can employ the
velocity direction as a training label. Although often slightly
incorrect, it remains correct on the average, to which our
point-estimator converges. Notably, the relation between velocity and
orientation holds regardless of the quality of the raw imaging data
employed. In case of overhead depth maps, as used here, often we had
disagreement between human annotators, which would possibly
unavoidably yield low quality labels. By using velocity we can
circumvent this issue and produce training data in arbitrarily large
amounts. It should also be stressed that this approach can be
conceptually extended to other imaging formats, such as color images,
provided accurate and sufficiently prolonged tracking data are
available.

Our tool unlocked the possibility to accurately investigate the
relation between velocity direction and shoulder orientation. We could
measure a velocity-dependent delay of about $100\,$ms between the
first and the second, that we are able to quantitatively reproduce in
terms of a simple Ornstein-Uhlenbeck process. In particular, on the
basis of measured orientation signals, we could generate velocity
directions featuring amplitude with respect to the orientation signal,
velocity-orientation delay distribution and power spectral density in
very good agreement with the measurements.

Our velocity-trained network could be possibly employed to investigate
conditions of static crowds, clogged bottlenecks conditions, or other
scenarios in which the ``nematic'' ordering of the crowd is expected
to play a key role in the dynamics.

\subsubsection*{Acknowledgments}
A.C. acknowledges the Talent Scheme (Veni) research programme (project
N. 16771) financed by the Netherlands Organization for Scientific
Research.

\bibliographystyle{ieeetr}
\bibliography{biblio} 

\newpage

\section*{Supplementary Information (SI)}

\subsection*{Arithmetics of angles on $\Pj^1(\R)$}
We choose to parametrize the projective line, $\Pj^1(\R)$, with the
interval $[-\frac{\pi}{2}, \frac{\pi}{2})$. An angular value,
$\theta' \in \R$, is reported to this parametrization of $\Pj^1(\R)$
through the $\wrap{}$ function, defined as
\begin{equation}
\wrap{(\theta')} = \bmod(\theta'+\pi/2, \pi) -\pi/2.
\label{eq:wrap}
\end{equation}
We define arithmetic operations such as angle summation and
subtraction of two angles, say $\theta_1, \theta_2 \in \Pj^1(\R)$, via
\eqref{eq:wrap} as
\begin{equation}
\theta_1 \pm \theta_2 = \wrap{(\theta_1 \pm \theta_2)}.
\label{eq:circdiff}
\end{equation}
Weighted averaging operations on $\Pj^1(\R)$ (e.g. \eqref{eq:Proj-avg}
and \eqref{eq:O2avg}) are computed in this parametrization as
\begin{equation} 
  \E_{\theta'\sim h(\theta), \Pj^1(\R)}[\theta']= \frac{1}{2} \arctantwo{\left( \int_{-\pi/2}^{\pi/2} h(\theta) \sin{(2\theta)} d\theta, \int_{-\pi/2}^{\pi/2} h(\theta) \cos{(2\theta)} d\theta \right)},
\label{eq:pdf_to_angle}
\end{equation}
where $h(\theta)$ is a probability density function on $\Pj^1(\R)$.
Namely, angles $\theta$ are converted to corresponding Cartesian
points on the unit circle (i.e.
$\theta \rightarrow (\cos{(2\theta)}, \sin{(2\theta)})$), then a vector
average weighted by $h(\theta)$ is performed, and the final result is
mapped back to $\Pj^1(\R)$ via $\frac{1}{2} \arctantwo(\cdot)$.  Note
that \eqref{eq:pdf_to_angle} is not defined whenever the vector
average vanishes. This happens, for instance, when $h(\theta)$ is
uniform.  For further details on arithmetic of periodic variables, we
refer to~\cite{jammalamadaka2001topics}.

\subsection*{``Two-hot'' encoding}
Our neural network outputs a discrete probability distribution over
$B=45$ classes. We interpret it as a probability over $[-\pi/2,\pi/2)$
once partitioned in $B$ equal adjacent intervals of size $\pi/B$ and
centered around the mid-value
$\theta_i = \pi(1/2B - 1) + i\,\pi/B  = -88^{\circ} + i\,4^{\circ}$,
with $i \in \{0, 1, 2,...,B-1\}$.
Let $h(\theta_i)$ be the considered probability distribution. In
support of the required circular properties, we enforce adjacency of
the outer bins, such that they ``wrap'' around $\Pj^1(\R)$, and
$h(\theta) = h(\theta + k\pi)$ for all $k \in \Z$ holds.

In the continuous case, $B\rightarrow \infty$, a ground truth
annotation (or the ideal output of the
neural network) is a Dirac delta probability distribution,
$\delta(x-\theta)$, centered on the true angle $\theta$. From this,
$\theta$ can be recovered via the expectation in
\eqref{eq:pdf_to_angle}:
$\theta = \E_{\theta\sim\delta(x-\theta)}[ \theta']$. For finite $B$,
we prevent quantization errors by unambiguously encoding angles in (up
to) two adjacent bins, from which the name ``two-hot'' encoding
(vs. the ``one-hot'' encoding in standard classification problems). In
particular we encode the angle $\theta$ as 
\begin{equation}
        h_2(\theta)[\theta_i] = 
        \mathcal{N}\begin{cases}
            1 - \frac{1}{d} \abs{\theta-\theta_i} &	\hspace{5mm} \text{if  }  \abs{\theta-\theta_i} \leq \frac{\pi}{B} \\
            0 					&	\hspace{5mm} \text{otherwise  } 
        \end{cases}
        \label{eq:angle_to_pdf}
\end{equation}
with $\mathcal{N}$ being a normalization constant and
$\theta-\theta_i$ being the wrapped distance (\eqref{eq:circdiff}) of
$\theta$ and the mid-angle of the $i$-th bin. Note that the wrapped
difference ensures that both $\theta = -\pi/2$ and $\theta = \pi/2$
are encoded into the same distributions, i.e. the network output
remains unchanged for a $180^{\circ}$ body rotation. By
applying~\eqref{eq:pdf_to_angle}, we recover the annotation angle
$\theta$.

\begin{figure}[h]
\centering
\includegraphics[width=0.5\textwidth]{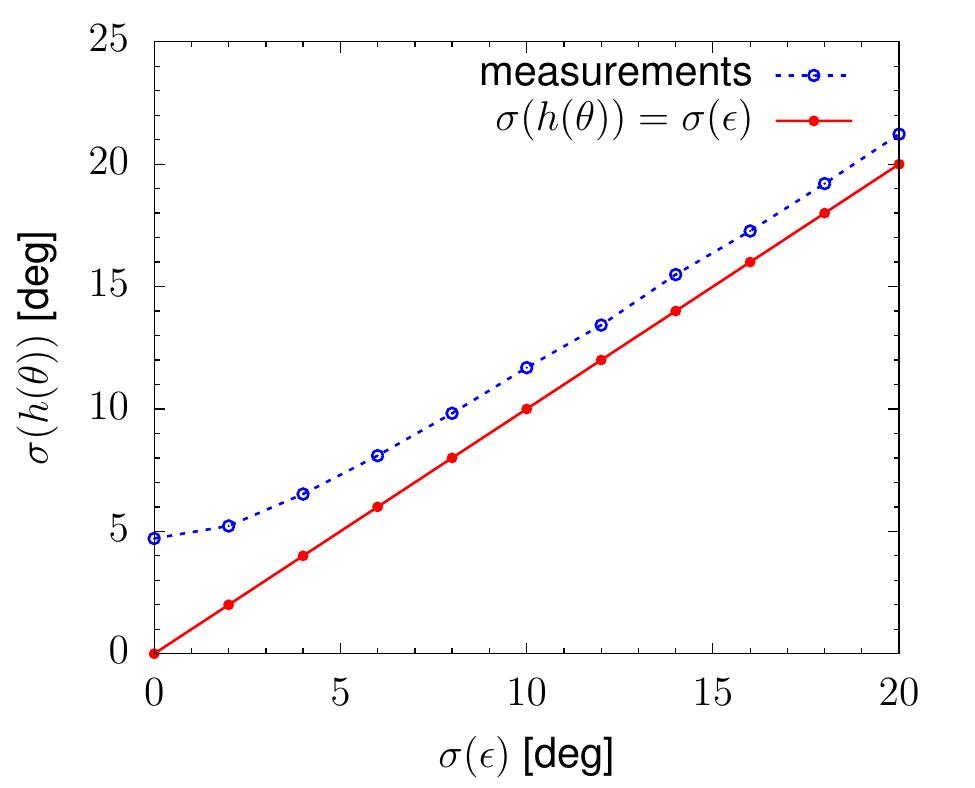}
\caption{Dispersion (standard deviation) in the neural network output
  probability distribution $\sigma(h_{pred})$ versus the amplitude
  (standard deviation) of the error term $\sigma(\epsilon)$ (between
  training annotations and ground-truth orientation in accordance
  with~\eqref{eq:velocity-proxy}) for synthetic imagelets. We train
  $M=16$ networks for each $\sigma(\epsilon) \in \{0, 2, 4,...,20\}$
  and evaluate $\sigma(h_{pred})$ by averaging the standard deviations
  of the outputs $h_{pred}$ among the $M$ networks on $10k$ synthetic
  test imagelets. These measurements are reported as blue dotted
  lines. 
  We observe good scaling agreement between $\sigma(h_{pred})$ and the
  theoretical lower limit $\sigma(\epsilon)$ (cf. proof in the text
  for the simplified case), indicating that our training converges to
  the average annotation for similar imagelets.}
\label{fig:noise-relation}
\end{figure}

\subsection*{Variance of $h(\theta)$ and cross-entropy based error
  averaging}
In this section we provide a physical meaning to the to the
probability distribution predicted by the neural network
$h_{\text{pred}}$, beyond the average value, which we retain as
predicted angle $\theta_o$. In particular, the standard deviation,
$\sigma(h_{\text{pred}})$, scales with the error $\epsilon$ of the
annotations given to similar imagelets.  We prove this claim formally
in a simplified case and via simulations in the general one.

Training by minimizing the average cross-entropy,
$\E_{\theta_v}[\loss(h_{\text{pred}},\delta_{\theta_v}))]$, in the
simplified scenario with the following characteristics: 1. $\Pj^1(\R)$
is considered in its discretized version, i.e. $B<\infty$; 2. there is
a single training imagelet, $\Il$, yet endowed with different
(conflicting) annotations; 3. annotations, $N$ in total, are
distributed as $\theta_v = \theta + \epsilon$ (where $\epsilon$ is
zero-averaged, see main text); 4. annotations are centered in some bin
$j\in\{0,1,\dots,B-1\}$ (i.e. they are Dirac masses,
$\delta_{\theta_j}$) yields
\begin{equation}
  h_{\text{pred}} = \hat h_{\theta+\epsilon},
\end{equation}
where $\hat h_{\theta+\epsilon}$ is the (discretized) probability
distribution of the training data. Notice that, by construction, the
$\Pj^1(\R)$-average of $h_{\text{pred}}$ is exactly $\theta$.

\noindent \textbf{Proof.} Let $h_{pred}$ have values $h(\theta_0),
h(\theta_1),\ldots,h(\theta_{B-1})$ on the $B$ bins.
The average loss, $\mathcal{L}$, reads
\begin{eqnarray}
\mathcal{L} &=& \frac{1}{N} \sum_{j=1}^N \loss(\delta_{\theta_{j}}, h_{pred}) \\ 
&=& -\frac{1}{N} \sum_{j=1}^N \sum_{i=0}^{B-1} \delta_{\theta_{j}} \log h(\theta_i) \\ 
&=& -\frac{1}{N} \sum_{j=1}^N \log h(\theta_\mu),\label{eq:unsorted-sum}
\end{eqnarray}
where the last equality follows from the definition of Dirac mass.

We can sort and aggregate the elements in the sum in
\eqref{eq:unsorted-sum}. In particular, let $\#_j$ be the total number
of annotations having value $\delta_{\theta_j}$, \eqref{eq:unsorted-sum} yields:
\begin{equation}
\mathcal{L} = -\sum_{j} \frac{\#_j}{N} \log h(\theta_j),
\end{equation}
by Gibbs' inequality (see, e.g.,~\cite{mackay2003information}), it holds 
\begin{equation}
-\sum_{j} \frac{\#_j}{N} \log h(\theta_j) \geq -\sum_{j} \frac{\#_j}{N} \log \frac{\#_j}{N}. \\ 
\end{equation}
Therefore, at the absolute minimum for $\mathcal{L}$, $h_{pred}$ is
the distribution of the labels, which in our case is  $\hat
h_{\theta+\epsilon}$, i.e.
\begin{equation}
h_{pred} = \sum_j \frac{\#_j}{N} \delta_{ \theta_j} = \hat h_{\theta+\epsilon} \label{eq : output_distribution}
\end{equation}\QED

\noindent Note that for a two-hot encoding the proof is identical, but each
annotation is a convex combination of two delta masses.

In general, in a sufficiently ample dataset of depth imagelets (and
related velocity annotation) acquired in absence of biases
compromising the relation~\eqref{eq:velocity-proxy}, we expect to find
a wide number of similar imagelets, yet with different annotations,
and this provides a rich sampling of the $\epsilon$
distribution. Abstracting from the previous proof, we expect the
training process to be such that for each set of similar imagelets,
the network would learn and output the probability distribution of
annotations. We prove this experimentally, by means of synthetic
imagelets. In Fig.~\ref{fig:noise-relation} we compare the amplitude
of the symmetric centered error, $\sigma(\epsilon)$, with the standard
deviation of the predicted distribution (averaging over a test set of
$10.000$ images), showing that they perfectly correlate for
$\sigma(\epsilon)>7^\circ$.

\subsection*{Neural network structure and training}\label{sect:NNet}
\begin{figure}[h]
\centering
\includegraphics[width=18cm]{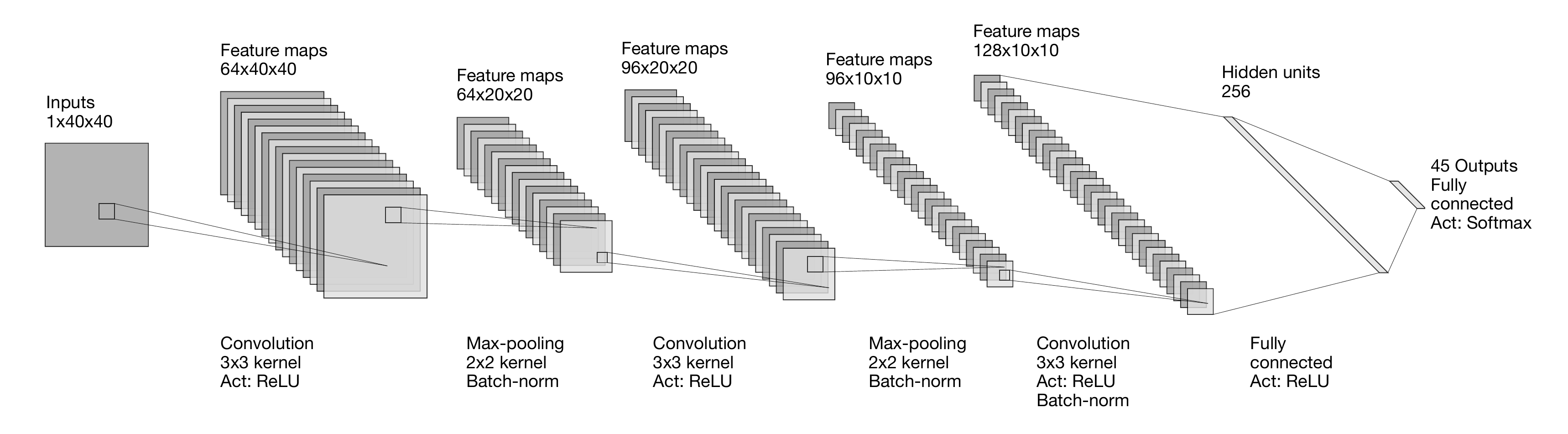}
\caption{Detailed structure of the neural network. The network is fed
  with single channel imagelets ($40\times40$ pixels) after which the
  input data propagates trough two stacks of convolution, max-pooling
  and batch normalization layers for feature extraction. A convolution
  and batch-normalization layer connects the feature maps with a fully
  connected layer (ReLU activation function). The final softmax
  activation yields a probability mass function $h(\theta)$ on $45$
  adjacent equal bins as output of the last layer. The network is
  trained using cross-entropy as loss
  function.\label{fig:NN-architecture}}
\end{figure}

We consider a neural network inspired by the VGG model, whose full
structure is in Figure~\ref{fig:NN-architecture}. We implemented the
network using the Keras library.

We train the network by randomly augmenting the training images at the
beginning of every epoch. Specifically, we apply random rotations and
random horizontal flips (and we act correspondingly to the associated
labels) to all imagelets.  This ensures a training dataset uniformly
distributed on $\Pj(\R^1)$. A pre-processing standardization step of
the depth intensity is applied individually to all the imagelets.

We employ the Adam optimizer with a batch size of $64$, we retain
the model that scores the lowest RMSE over a total number of $25$
training epochs.

\newpage

\subsection*{Map $\tilde f$ and $O(2)$-group averaging}
In this Section, we deduce identity~\eqref{eq:O2avg}, and prove that
the map $\tilde{f}$ is strictly respecting $O(2)$ symmetry
(cf.~\eqref{eq:f-sym}). 

The identity between \eqref{eq:O2avg} and \eqref{eq:Otwo-avg} can be
proved by substitution, considering the fact that $O(2)$ can be
decomposed into rigid rotations and rigid rotations applied after a
reflection, from which \eqref{eq:decompSO2}:
\begin{eqnarray}
\tilde{f}(\Il) &=& \frac{1}{|O(2)|}\int_{O(2)}(f(\phi \Il) - \alpha)\det(\phi)\, d\phi\\
&=& \frac{1}{2} \left( \underbrace{\frac{1}{2\pi}
    \int_{0}^{2\pi}f(R_{\alpha} \Il) - \alpha
    d\alpha}_{\text{rotations:} \det(\phi)=1} +
    \underbrace{\frac{1}{2\pi} \int_{0}^{2\pi}-f(J R_{\alpha} \Il) -
    \alpha d\alpha}_{\text{reflection and rotations:} \det(\phi)=-1} \right) \label{eq:decompSO2}\\ 
&=& \frac{1}{2\pi} \int_{0}^{2\pi} \frac{1}{2} \left( \left\{ f(R_{\alpha} \Il) - f(J R_{\alpha} \Il) \right\} - 2\alpha \right) d\alpha \\ 
&=&\frac{1}{2\pi}\int_{0}^{2\pi}\frac{f(R_\alpha\Il) - f(JR_\alpha \Il)}{2} - \alpha\, d\alpha 
\end{eqnarray} \QED \\

\noindent We show here that $\tilde{f}$ respects $O(2)$ symmetry
(\eqref{eq:f-sym}).

\noindent \textbf{Proof.} Let $\phi \Il=\Phi R_{\beta} \Il$ be a
rotation of $\beta$ and, possibly, a reflection (i.e. $\Phi \in
\{\text{Id}, J\}$) applied to $\Il$. From the \eqref{eq:Otwo-avg}, $\tilde{f}$ reads
\begin{equation}
\tilde{f}(\phi \Il) =\frac{1}{2\pi}\int_{0}^{2\pi}\frac{f(R_\alpha \phi \Il) - f(JR_\alpha \phi \Il)}{2} - \alpha\, d\alpha.
\end{equation}
We prove that $\tilde{f}$ respects \eqref{eq:f-sym} by addressing the
cases $\Phi=\text{Id}$ ($\phi$ does not include a reflection) and
$\Phi=J$ ($\phi$ does include a reflection).

\noindent \textit{Case} - $\det{(\Phi)}=1$, i.e.
$\tilde{f}(\phi \Il) = \tilde{f}(R_{\beta} Id \Il) =
\tilde{f}(R_{\beta} \Il)$.

\noindent By definition of $\tilde{f}$, the group average yields 
\begin{eqnarray}
\tilde{f}(\phi \Il)&=&\frac{1}{2\pi}\int_{0}^{2\pi}\frac{f(R_{\alpha}R_{\beta}\Il) - f(J R_{\alpha} R_{\beta} \Il)}{2} - \alpha\, d\alpha \\ 
&=&\frac{1}{2\pi}\int_{0}^{2\pi}\frac{f(R_{\alpha+\beta}\Il) - f(J R_{\alpha+\beta} \Il)}{2} - \alpha\, d\alpha,
\end{eqnarray}
which, after applying the rotation of $\alpha+\beta = \gamma$, becomes
\begin{eqnarray}
&=&\frac{1}{2\pi} \left( \int_{0}^{2\pi}\frac{f(R_{\gamma}\Il) - f(J R_{\gamma} \Il)}{2} - \gamma d\gamma, + \int_{0}^{2\pi} \beta\, d\beta \right) \\ 
&=&\frac{1}{2\pi}  \int_{0}^{2\pi}\frac{f(R_{\gamma}\Il) - f(J R_{\gamma} \Il)}{2} - \gamma\, d\gamma + \beta \\ 
&=& \tilde{f}(\Il) + \beta \label{eq:det+1}
\end{eqnarray}

\noindent \textit{Case 2} - $\det{(\Phi)}=-1$, i.e. $\tilde{f}(\phi \Il) = \tilde{f}(R_{\beta} J \Il)$ 

\noindent The group average $\tilde{f}$ yields 
\begin{equation}
\tilde{f}(\phi \Il) = \frac{1}{2\pi}\int_{0}^{2\pi}\frac{f(R_{\alpha} J R_{\beta}  \Il) - f(J R_\alpha J R_{\beta} \Il)}{2} - \alpha\, d\alpha. 
\end{equation}
The order in which mirroring and rotation are applied determines the
sign of the rotation. For an angle $\theta$, the identity
$R_{\alpha}J\theta = (\pi - \theta)+\alpha = \pi - (\theta - \alpha) =
J R_{-\alpha} \theta$ holds. By using this fact, we get
\begin{eqnarray}
&=& \frac{1}{2\pi}\int_{0}^{2\pi}\frac{f(R_{-\alpha} R_{\beta} J  \Il) - f(J J R_{-\alpha}  R_{\beta} \Il)}{2} - \alpha\, d\alpha \\ 
&=& \frac{1}{2\pi}\int_{0}^{2\pi}\frac{f(R_{-\alpha+\beta} J \Il) - f(R_{-\alpha+\beta} \Il)}{2} - \alpha\, d\alpha \\ 
&=& \frac{1}{2\pi}\int_{0}^{2\pi} - \frac{f(R_{-\alpha+\beta} \Il) - f(J R_{-\alpha+\beta} \Il)}{2} - \alpha\, d\alpha, 
\end{eqnarray}
which, after applying the transformation $-\alpha+\beta = \gamma$, becomes
\begin{eqnarray}
&=& \frac{1}{2\pi}\int_{\beta}^{-2\pi+\beta} - \frac{f(R_{\gamma} \Il) - f(J R_{\gamma} \Il)}{2} + \gamma - \beta (-d\gamma) \\ 
&=& \frac{1}{2\pi}\int_{0}^{-2\pi}  \frac{f(R_{\gamma} \Il) - f(J R_{\gamma} \Il)}{2} - \gamma + \beta d\gamma \\ 
&=& \frac{1}{2\pi}\int_{0}^{-2\pi} \frac{f(R_{\gamma} \Il) - f(J R_{\gamma} \Il)}{2} - \gamma + \beta d\gamma \\
&=& \frac{-1}{2\pi}\int_{0}^{2\pi} \frac{f(R_{\gamma} \Il) - f(J R_{\gamma} \Il)}{2} - \gamma d\gamma + \beta \\
&=& -(\tilde{f}(\Il) + \beta). \label{eq:det-1}
\end{eqnarray}
Hence, by combining ~\eqref{eq:det+1} and~\eqref{eq:det-1}, the
proposition holds
\begin{equation}
 \tilde{f}(\phi \Il) = (\tilde{f}(\Il) + \beta)\det(\phi)
\end{equation}\QED

We discretize the otherwise continuous $O(2)$-averaging by considering
an equi-spaced sampling of the circle and a random sampling, whose
results are reported in Fig.~\ref{fig:examples-of-tracks}.

\subsection*{Time delay of two signals}

Let $\theta(\omega)$ and $\theta_v(\omega)$ be the Fourier transform
of the signals $\theta(t)$ and $\theta_v(t)$, respectively. By
applying the argument operator, $\arg{(\cdot)}$, we can compute the
corresponding phase as function of the frequency,
i.e. $\phi_{\theta}(\omega)$ and $\phi_{\theta_v}(\omega)$. This
enables to compute the delay time for each frequency component:
\begin{equation}
\tau(\omega) = \frac{\phi_{\theta}(\omega) - \phi_{\theta_v}(\omega)}{2\pi}.
\end{equation}
We retain as characteristic delay time between the signals the value
$\tau(\omega^*)$, provided $\omega^*$ exists, according to the
following procedure:
considering the frequency range $f \in [0.6, 1.2]\,$Hz, where the
typical walking fluctuations occur, we compute 
\begin{eqnarray}
  \omega_{\theta} &=& \argmax{_{f \in [0.6, 1.2] \text{Hz}} \big( E_{\theta} \big)} \\ 
  \omega_{\theta_v} &=& \argmax{_{f \in [0.6, 1.2] \text{Hz}} \big( E_{\theta_v} \big)} 
\end{eqnarray}
where $E_{\theta}$ and $E_{\theta_v}$ are the energy spectra that can be
computed by application of the module operator $\abs{\cdot}$ to the
 $\theta(\omega)$ and $\theta_v(\omega)$. We set $\omega^*= \omega_{\theta}$
 if $\omega_{\theta}\approx\omega_{\theta_v}$, i.e. velocity and
 orientation are synchronized. Else, we discard the trajectory from the
 computation of the delay.

\subsection*{Generation of synthetic data} % \label{sec:train_synthetic}
We generate real-life imagelets mimicking the overhead shape of people
in terms of a superposition of two ellipses: one for the
body/shoulder, $E_b$, and another one, $E_h$, at lower depth values
(larger height), for the head
(cf. Figure~\ref{fig:synthetic_examples}(a,b)). We characterize each
ellipse, by $6$ random scalars: Cartesian coordinates $x_j, y_j$ of
its center, area $A_j$, eccentricity $e_j$, rotation angle $\alpha_j$
and depth value $c_j$ (i.e. the gray shade in the colorization in
Fig.~\ref{fig:challenges}, $j=b,\,h$). A bivariate normal random
distribution around the imagelet center determines the position of the
body ellipse. The head ellipse is superimposed at a uniform random
position closely to the end of the body ellipse minor axis.  We add
exceptions that are often seen in real-life data (i.e. perturbations
of the overhead elliptical pedestrian shape) by drawing $4$ additional
ellipses at random positions in the imagelets. Moreover, artifacts of
adjacent pedestrians in imagelets due to local dense situations are
represented in the dataset by cropping $9$ imagelets from a
$3 \times 3$ grid in which pedestrians are drawn at random positions
relative to these grid
points. 
We represent clothing artifacts, shape variations and imperfect depth
reconstruction by applying random modifications to each synthetic
imagelet (in the same spirit of~\cite{corbetta2017weakly}): these
include pixel removal (i.e. replacing $15\%$ random pixels with
background pixel value), pixel addition (i.e. replacing $25\%$ with
median pixel value), depth translation (i.e. increasing the values of
all foreground pixels with a single uniform random variable
$x_i \in [-15, 15]$) and Gaussian noise ($\sigma=5$). Finally, we
smoothen by convolving the imagelets with a $3\times3$ averaging
kernel, resulting in the synthetic imagelets of
Figure~\ref{fig:synthetic_examples}. We report the imagelet generation
algorithm in full in Algorithm~\ref{algo:image}.

\begin{algorithm}[b!]
\caption{Algorithm for the generation of synthetic imagelets (cf. Fig.~\ref{fig:synthetic_examples}). Each iteration of the nested for-loop draws a single pedestrian on a square background $\beta$ at relative distance $d$ to create artifacts of adjacent pedestrians. Random variables (generated in lines $6-19$) characterise two ellipses (drawn in lines $20$ and $21$) that represent the body and head. We introduce a $25\%$ probability of drawing children, represented by smaller ellipses (lines $18$ and $19$). Additionally, perturbations of the elliptical overhead shape (e.g. due to backpacks, arms or posture) are imitated by drawing $N_{obj}=4$ small ellipses at random positions in lines $23-29$. Finally, $9$ imagelets are obtained by cropping around each of the $3\times3$ grid positions (lines $30$-$36$).}\label{algo:image}
\begin{flushleft}
        \textbf{INPUT:} $N_{obj}=4, d=35$\\
        \textbf{OUTPUT:} $9$ imagelets $\mathcal{I}$ and $9$ labels $\theta_{gt}$
\end{flushleft}
	\begin{algorithmic}[1]
	
	\State{$\beta$ := square background with $(2d + 40)^2$ pixels and pixel values $255$}
	
	\State{$\mathcal{I}$ := object containing imagelets with $(40\times40)$ resolution}
	
	\State{$\theta_{gt}$ := object containing ground-truth labels}
	
	\item[]

      \For{$i=0$ to $3$}
        \For{$j=0$ to $3$}
        
    		\item[]    	
	
        	\State{$\vec{r}_{grid} := \big((35 + d i + \eta_1 )\cdot \vec{e}_x ,  (35 + d j + \eta_2)\cdot \vec{e}_y \big)$ with $\eta_1, \eta_2 \sim \mathcal{N}\left(0, 2\right)$}
		
			\State{$x_{h} := r_{grid,x} + \eta $ with $\eta \sim \mathcal{N}\left(0, 2\right)$ }	
			
			\State{$y_{h} := r_{grid,y} + \mu $ with $\mu \sim \mathcal{U}\left(0, 7\right)$ }
			
			\State{$\epsilon_{b}, \epsilon_{h} := (\eta_1, \eta_2)$ with $\eta_1, \eta_2 \sim \mathcal{N}\left(1.6, 0,2\right)$}
			
			\While{$\epsilon_{b} < 1.35$ or $\epsilon_{h} < 1.0$}
				\State{$\epsilon_{b}, \epsilon_{h} := (\eta_1, \eta_2)$ with $\eta_1, \eta_2 \sim \mathcal{N}\left(1.6, 0,2\right)$}
			\EndWhile
				
			\State{$A_{b} :=  \mu $ with $\eta \sim \mathcal{N}\left(600, 80\right)$}
			\State{$A_{h} := A_b \cdot \eta$ with $\eta \sim \mathcal{N}\left(0.3, 0.03 \right)$}
 
			\State{$c_b :=  \eta $ with $\eta \sim \mathcal{N}\left(170, 12\right) $}
			
			\State{$c_h :=  \eta $ with $\eta \sim \mathcal{N}\left(155, 8\right) $}
			
			\State{$\alpha_b := -\frac{\pi}{2} + \mu $ with $\mu \sim \mathcal{U}\left(0, \pi \right)$}
			
			\State{$\alpha_h := \alpha_b + \eta $ with $\eta \sim \mathcal{N}\left(-\frac{\pi}{6}, \frac{\pi}{6} \right)$}
			
			\While{$\mu > 0.75$ with $\mu \sim \mathcal{U}\left(0, 1\right)$}
				
					\State{Reduce $A_b, A_h, c_b, c_h$ with $25\%$}
					
			\EndWhile
			
			\item[]
			
			\State{$\beta$.DrawEllipse$\big(\vec{r}_{grid}, A_b, \epsilon_b, \alpha_b, c_b,  \big)$}
		
			\State{$\beta$.DrawEllipse $\big((x_h \vec{e}_x, y_h \vec{e}_y), A_h, \epsilon_h, \alpha_h, c_h,  \big)$}
			
			\State{$\theta_{gt}$.append$\big(\alpha_b\big)$}
			\item[]
			
			\For{$k=0$ to $N_{obj}$}
			
				\State{$\vec{r}_{obj} := \big(\eta\cdot \vec{e}_x ,  \eta\cdot \vec{e}_y \big)$ with $\eta \sim \mathcal{N}\left(10, 4\right)$}
		
				\State{$\epsilon_{obj} := \mu $ with $\mu \sim \mathcal{U}\left(1, 2\right)$}
				
				\State{$A_{obj} :=  \mu $ with $\eta \sim \mathcal{N}\left(100, 12\right)$}
					
				\State{$c_{obj} :=  \eta $ with $\eta \sim \mathcal{N}\left(170, 10\right) $}
				
				\State{$\alpha_{obj} := -\frac{\pi}{2} + \mu $ with $\mu \sim \mathcal{U}\left(0, \pi \right)$}
			
				\State{$\beta$.DrawEllipse $\big(\vec{r}_{obj}, A_{obj}, \epsilon_{obj}, \alpha_{obj}, c_{obj},  \big)$}
				
			\EndFor
			
			\item[]
			
	\EndFor
     		 \EndFor

			\For{$m=0$ to $3$}
				
				\For{$n=0$ to $3$}
				
					\State{$\vec{r}_{crop} := (\big((35 + m d)\cdot \vec{e}_x ,  (35 + n d + \eta)\cdot \vec{e}_y \big)$}
					
					\State{img := $\beta$.Crop$\big($ at = $\vec{r}_{crop}\big)$}
					
					\State{img := img.ApplyNoise$\big(\big)$}
					
					\State{img := img.Smoothen$\big(\text{kernel}=(3\times3)\big)$}
									
					\State{$\mathcal{I}$.append$\big($ img $\big)$}
			
				\EndFor
			
			\EndFor
			
			\item[]
      
      \Return{$\mathcal{I}, \theta_{gt}$}

	\end{algorithmic}
\end{algorithm}

\end{document}